\documentclass[fleqn,usenatbib]{mnras}

\usepackage{newtxtext,newtxmath}

\usepackage[T1]{fontenc}

\DeclareRobustCommand{\VAN}[3]{#2}
\let\VANthebibliography\thebibliography
\def\thebibliography{\DeclareRobustCommand{\VAN}[3]{##3}\VANthebibliography}

\usepackage{graphicx}	
\usepackage{amsmath}	
\usepackage{xcolor}

\title[Tidal Features in Cosmological Simulations]{Characterising Tidal Features Around Galaxies in Cosmological Simulations}

\author[A. Khalid et al.]{
A. Khalid,$^{1,2}$\thanks{E-mail: aman.khalid@unsw.edu.au}
S. Brough,$^{1,2}$
G. Martin,$^{3,4,5}$
L. Kimmig,$^{6}$
C. D. P. Lagos,$^{2,7,8}$
R.-S. Remus,$^{6}$
\newauthor
C. Martinez-Lombilla,$^{1,2,9}$
\\
$^{1}$School of Physics, University of New South Wales, NSW 2052, Australia\\
$^{2}$ARC Centre of Excellence for All Sky Astrophysics in
3 Dimensions (ASTRO 3D)\\
$^{3}$School of Physics and Astronomy, University of Nottingham, University Park, Nottingham NG7 2RD, UK\\
$^{4}$Korea Astronomy and Space Science Institute, 776 Daedeokdae-ro, Yuseong-gu, Daejeon 34055, Korea\\
$^{5}$ Steward Observatory, University of Arizona, 933 N. Cherry Ave, Tucson, AZ 85719, USA\\
$^{6}$Universitäts-Sternwarte München, Fakultät für Physik, LMU München, Scheinerstr. 1, D-81679 München,
Germany\\
$^{7}$International Centre for
Radio Astronomy Research, The University of Western Australia, 35 Stirling Highway, Crawley, WA 6009, Australia\\
$^{8}$Cosmic Dawn Center (DAWN)\\
$^{9}$Centre for Astrophysics and Supercomputing, Swinburne University of Technology, Hawthorn, VIC 3122, Australia
}

\date{Accepted XXX. Received YYY; in original form ZZZ}

\pubyear{2023}

\begin{document}
\label{firstpage}
\pagerange{\pageref{firstpage}--\pageref{lastpage}}
\maketitle

\begin{abstract}
Tidal features provide signatures of recent mergers and offer a unique insight into the assembly history of galaxies. The Vera C. Rubin Observatory’s Legacy Survey of Space and Time (LSST) will enable an unprecedentedly large survey of tidal features around millions of galaxies. To decipher the contributions of mergers to galaxy evolution it will be necessary to compare the observed tidal features with theoretical predictions. Therefore, we use cosmological hydrodynamical simulations \textsc{NewHorizon}, \textsc{EAGLE}, \textsc{IllustrisTNG}, and \textsc{Magneticum} to produce LSST-like mock images of $z\sim0$ galaxies ($z\sim0.2$ for \textsc{NewHorizon}) with $M_{\scriptstyle\star,\text{ 30 pkpc}}\geq10^{9.5}$ M$_{\scriptstyle\odot}$. We perform a visual classification to identify tidal features and classify their morphology. We find broadly good agreement between the simulations regarding their overall tidal feature fractions: $f_{\textsc{NewHorizon}}=0.40\pm0.06$, $f_{\textsc{EAGLE}}=0.37\pm0.01$, $f_{\textsc{TNG}}=0.32\pm0.01$ and $f_{\textsc{Magneticum}}=0.32\pm0.01$, and their specific tidal feature fractions. Furthermore, we find excellent agreement regarding the trends of tidal feature fraction with stellar and halo mass. All simulations agree in predicting that the majority of central galaxies of groups and clusters exhibit at least one tidal feature, while the satellite members rarely show such features. This agreement suggests that gravity is the primary driver of the occurrence of visually-identifiable tidal features in cosmological simulations, rather than subgrid physics or hydrodynamics. All predictions can be verified directly with LSST observations. 
\end{abstract}




\section{Introduction}

\begin{figure*}
    \centering
    \includegraphics[width=\linewidth]{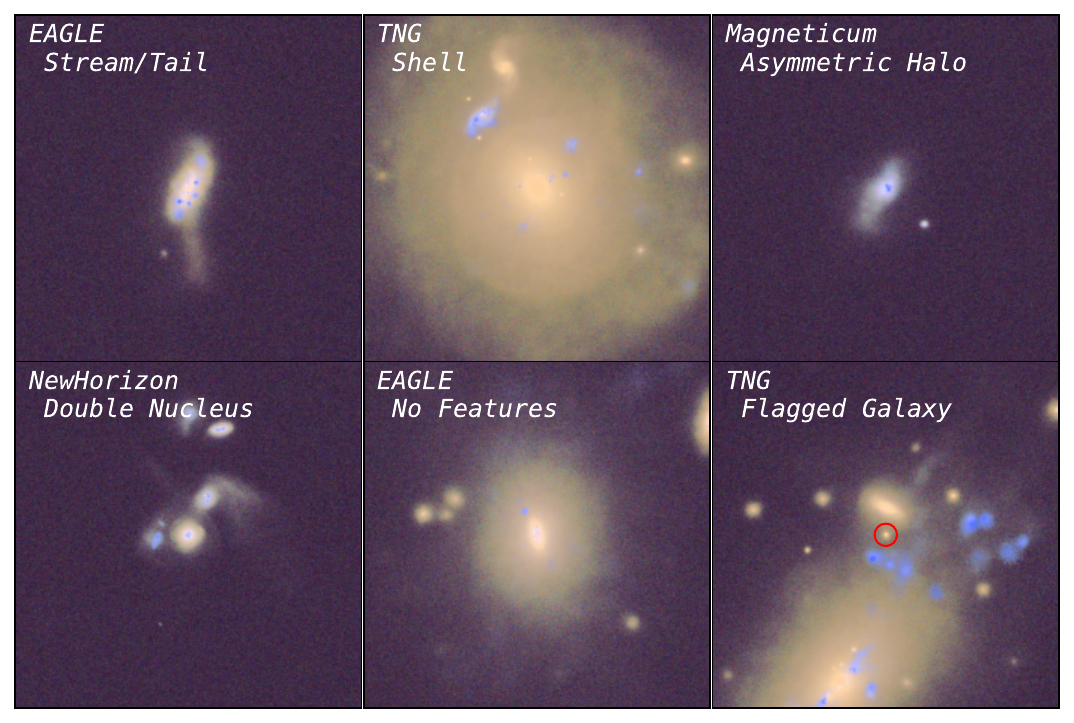}
    \caption{Examples of \textit{gri}-colour mock images from each simulation in our sample. Each of the tidal feature morphologies are illustrated from left to right, top to bottom: stream/tail, shell, asymmetric halo, double nuclei and featureless. The galaxy marked by a red circle in the bottom right image is an example of one that we removed from the sample. While there is enough stellar mass within the 30 pkpc aperture for the object to be in our sample, the object itself is too dense and compact to be an accurate representation of a galaxy of $M_{\scriptstyle\star,\text{30 pkpcs}}\geq10^{9.5}$ M$_{\scriptstyle\odot}$. Each image is 124 pkpc $\times$ 124 pkpc ($\sim2400\text{ pixels}\times2400\text{ pixels}$).}
    \label{fig:mock_images_tidal_features}
\end{figure*}

In the `hierarchical structure formation' model of the Universe \citep[e.g.][]{pressFormationGalaxiesClusters1974, fallFormationRotationDisc1980, rydenGALAXYFORMATIONGRAVITATIONAL1987, vandenboschUniversalMassAccretion2002, agertzFormationDiscGalaxies2011}, mergers play an important role in the evolution of galaxies, transforming galaxies through a variety of non-secular processes e.g. merger-driven star formation, merger-enhanced active galactic nuclei activity, dynamical evolution and stellar mass accretion \citep[e.g.][]{duboisHORIZONAGNSimulationMorphological2016,martinRoleMergersDriving2018,davisonEAGLEViewEx2020,martinRoleMergersInteractions2021,remusAccretedNotAccreted2022,cannarozzoContributionSituEx2023}. However, the intricacies of this picture, regarding the detailed merger statistics of galaxies, remain unresolved. There remain untested theoretical predictions from cosmological simulations. For example, \citet{naabAreDiskGalaxies2009} and \citet{martinRoleMergersDriving2018} predicted that low-mass mergers are the major contributor to galaxy morphological evolution since $z\sim1$. There are also detailed predictions regarding the environmental dependence of galaxy interaction and merger rates from cosmological-N-body simulations \citep[e.g.][]{gnedinTidalEffectsClusters2003, mihosInteractionsMergersCluster2003} and semi-analytic models \citep{jianEnvironmentalDependenceGalaxy2012}, where the galaxy encounter rates increase with increasing halo mass to a peak at halo masses corresponding to galaxy groups before dropping off at halo masses above that, corresponding to large galaxy clusters. To test these predictions we need to better understand galaxy mergers.

There are several ways to measure galaxy mergers observationally. These include detecting visible signatures of ongoing and past mergers via the presence of tidal features and selecting close pairs of galaxies that are likely to merge in the near future. Tidal features are diffuse non-uniform regions of stars that extend out from a galaxy, signatures of proceeding and concluded mergers in the forms of `tails', `streams', `asymmetric halos', `double nuclei' and `shells' (examples in Figure \ref{fig:mock_images_tidal_features}). Observations and numerical simulations have found these features to have lifetimes of $\sim3$ Gyrs \citep{jiLifetimeMergerFeatures2014,mancillasProbingMergerHistory2019,yoonFrequencyTidalFeatures2020,huangMassiveEarlyTypeGalaxies2022}, making tidal features crucial probes for a galaxy's recent merger history. Tidal feature detection requires very deep images to pick up signatures of ongoing and past mergers \citep[e.g.][]{atkinsonFAINTTIDALFEATURES2013,kado-fongTidalFeatures052018, martinPreparingLowSurface2022}. Close pairs are detected most accurately using spectroscopic observations to measure robust 3D positions for each galaxy \citep[e.g.][]{robothamGalaxyMassAssembly2014,banksGalaxyMassAssembly2021}. Close pair detection is important in understanding the role mergers play in driving galaxy evolution, however, this method does have limitations. It cannot detect the presence of a secondary galaxy if: that galaxy has been ripped apart by tidal forces; absorbed into a host galaxy; or the secondary galaxy is too low mass to detect via spectroscopy \citep[e.g.][]{lotzMAJORMINORGALAXY2011,desmonsGalaxyMassAssembly2023}. Studying the tidal features of a galaxy provides insight into these aspects of the merging process. 

The formation pathways of tidal features have been studied using N-body and hydrodynamic simulations to obtain insight into the merger histories of observed galaxies that host tidal features. These simulations have shown that tidal features are tracers of galaxy interactions and mergers. They probe detailed properties of these interactions (e.g. dynamics, mass ratios and orbits). They have established that tail-like structures (top left panel Figure \ref{fig:mock_images_tidal_features}) are formed from high angular momentum passages between similar mass galaxies (common in major mergers), while streams form from almost circularly infalling lower mass satellite galaxies \citep[minor mergers; e.g.][]{toomreGalacticBridgesTails1972, hendelTidalDebrisMorphology2015, karademirOuterStellarHalos2019}.

Various formation scenarios have been proposed to explain the formation of shells, the scenario most supported by analytical and numerical models is formation predominantly through radial mergers \citep[e.g.][]{hendelTidalDebrisMorphology2015,amoriscoFeathersBifurcationsShells2015, popFormationIncidenceShell2018,karademirOuterStellarHalos2019, valenzuelaStreamComeTrue2023}, resulting in overdensities of stripped stars accumulating at the apocentres of their orbits. Simulations have linked the properties of shells to the merger dynamics and mass ratios of the progenitor galaxies. N-body studies such as \citet{hernquistFormationShellsMajor1992} have found that shells formed from major mergers tend to be preferentially aligned with the major axis of the host galaxy. \citet{popFormationIncidenceShell2018} found that most of the $z=0$ shell population of the hydrodynamic-cosmological simulation \textsc{Illustris} were formed predominantly through major mergers, suggesting that due to dynamical friction, pairs of more massive galaxies could probe a greater range of impact parameters (deviating further from a purely radial infall) and still form shells when compared to more minor mergers. These simulations provide strong evidence for shell formation occurring through mergers with more radial trajectories than those that form tails and streams.

The understanding of tidal feature formation from simulations is used to interpret observational studies of the characteristics of tidal features such as their colours, dynamics, appearance and the environment in which the tidal feature host resides. These are then used to constrain the properties of the interaction that caused them. For example \citet{fosterKinematicsSimulationsStellar2014} and \citet{martinez-delgadoFeatherHatTracing2021} used spectroscopic and photometric data, respectively, in conjunction with numerical simulations, to study the formation of tidal features around the NGC 4651 \citep{fosterKinematicsSimulationsStellar2014} and M104 \citep{martinez-delgadoFeatherHatTracing2021} galaxies and constrain their recent merger histories. \citet{fosterKinematicsSimulationsStellar2014} were able to infer the recent passage of a satellite galaxy through the disk of NGC 4651 from their modelling of an observed stellar stream. \citet{martinez-delgadoFeatherHatTracing2021} were able to estimate the time around which the major wet merger forming M104 occurred. These particular examples illustrate the strength of tidal features in reconstructing merger histories of observed galaxies and hint at the potential of using tidal features to analyse larger statistical populations of galaxies to provide a detailed understanding of the merger process in our Universe.

One of the largest observational studies of tidal features to date, \citet{kado-fongTidalFeatures052018}, analysed the tidal features in $0.05<z<0.45$ galaxies in the HSC-SSP (Hyper Suprime-Cam Subaru Strategic Program) Wide layer. They found evidence suggesting that shells are formed predominantly through minor mergers.

Cosmological hydrodynamical simulations enable the study of large statistically-significant samples of galaxies to predict the dominant pathways of merger signatures in the $\Lambda$CDM model of hierarchical structure formation. Observations provide us with snapshots of our Universe at various lookback times, giving us the ground truth to which we can compare our models. Studying both observations and simulations enables us to understand the merger histories that are inferred from observations.

With the upcoming Vera C. Rubin Observatory's Legacy Survey of Space and Time \citep[LSST;][]{ivezicLSSTScienceDrivers2019a,robertsonGalaxyFormationEvolution2019,broughVeraRubinObservatory2020} it will be possible to study tidal features around millions of galaxies \citep{martinPreparingLowSurface2022}, allowing for the most robust statistical survey of tidal features to date. To fully capitalise on this unprecedented wealth of data we will need robust predictions of the properties of these features from current hydrodynamic-cosmological simulations. In this study, we will characterise the tidal features around galaxies in the four hydrodynamic-cosmological simulations (described in \S\ref{subsec:simulations}), \textsc{NewHorizon} \citep{duboisIntroducingNEWHORIZONSimulation2021}, \textsc{IllustrisTNG} \citep{springelFirstResultsIllustrisTNG2018, nelsonFirstResultsIllustrisTNG2018, marinacciFirstResultsIllustrisTNG2018, naimanFirstResultsIllustrisTNG2018}, \textsc{EAGLE} \citep{schayeEAGLEProjectSimulating2015, crainEAGLESimulationsGalaxy2015} and \textsc{Magneticum} \citep{tekluConnectingAngularMomentum2015}, in an observationally-motivated approach that facilitates future comparison to LSST. We do this by visually classifying the tidal features present in LSST-like mock images, produced following \citet{martinPreparingLowSurface2022}. We study four simulations to probe whether the different subgrid physics models applied to each simulation result in different merger pathways and whether this presents differently in the tidal features of galaxies. We compare the limited observations made to date to test how well the simulations are able to replicate them.

Our sample selection and production of mock images are described in \S\ref{subsec:sample_selection} and \S\ref{subsec:mock_images}, respectively. Our methodology for visual classification, including the tidal feature morphologies considered, are given in \S\ref{subsec:tidal_feature_classification} and we describe the methodology for calculating the errors on our tidal feature fractions in \S\ref{subsec:caculating_errors}. We present a statistical analysis of our visual classifications, considering the occurrence rates of tidal features in each simulation in \S\ref{sec:results} and then present how the fraction of galaxies exhibiting tidal features changes with both stellar mass (in \S\ref{subsec:results_mstar}) and halo mass (in \S\ref{subsec:results_m200}). We then discuss our results with respect to past analyses of simulations and observational results in \S\ref{subsec:simulations_comparison} and \S\ref{subsec:observations} respectively. The implications of our visual classification method are addressed and discussed in \S\ref{subsec:discussion_visual_classification}. In \S\ref{subsec:differences} we consider the implications of our results on the occurrence and observability of tidal features across the cosmological hydrodynamical simulations that we study. We discuss the relationship between tidal feature occurrence in the hydrodynamic-cosmological simulations we study and the environment in which the galaxies reside in \S\ref{subsec:environment}. In \S\ref{subsec:lsst_predictions} we explore the predictions for LSST regarding tidal feature fractions and their trends with stellar and halo mass based on the hydrodynamic-cosmological simulations studied here. We draw our conclusion in \S\ref{sec:conclusions}.

We use the native cosmology from each simulation for calculating the distances between particles and creating our mock images, these are given in Table \ref{tab:simulations}. For distances we use a `c' prefix to denote comoving coordinates and a `p' prefix to denote proper coordinates, e.g. cMpc is comoving megaparsecs and pMpc is proper megaparsecs.

\section{Simulations and methods}
\begin{table*}
    \centering
    \caption{Summary of the properties of the four cosmological hydrodynamical simulations. From left to right, the columns are the simulation name, matter density, dark energy density, Hubble constant, simulation box side length, stellar mass resolution, the redshift corresponding to the simulation timestep at which the particle data used was recorded, the number of galaxies in the simulations with $M_{\scriptstyle\star,\text{ 30 pkpc}}\geq10^{9.5}$M$_{\scriptstyle \odot}$, the number of galaxies classified, number of compact point-like galaxies discarded, and final sample size for analysis.}
    \begin{tabular}{|p{3.0cm}|p{0.8cm}|p{0.8cm}|p{1.8cm}|p{0.8cm}|p{1.cm}|p{1.4cm}|p{0.8cm}|p{0.8cm}|p{0.8cm}|p{ .8cm}}
        \hline
        Simulation & $\Omega_m$ & $\Omega_\Lambda$ & $H_0$ \newline[km s$^{-1}$Mpc$^{-1}$] & $V_{\rm box}$ \newline[$\text{cMpc}^3$] & $m_{\scriptstyle\star}$ \newline$[\text{ M}_{\scriptstyle \odot}]$ & $z$ & $N_{\text{gals}}$ & $N_{\text{classified}}$ & $N_{\text{out}}$ & $N_{\text{sample}}$\\
        \hline
        \textsc{NewHorizon} & 0.272 & 0.728 & 70.4 &  $16^3$ & $1.3\times10^4$ & 0.260, 0.263 & 105 & 62 & 0 & 62\\
        \textsc{EAGLE RefL0100N1504} & 0.307 & 0.693 & 67.8 & $100^3$ & $1.8\times10^6$ & 0.05 & 7273 & 2000 & 22 & 1978\\
        \textsc{TNG L75n1820TNG} & 0.309 & 0.691 & 67.7 & $111^3$ & $1.4\times10^6$ & 0.05 & 12418 & 1907 & 81 & 1826\\
        \textsc{Magneticum Pathfinder Box4-uhr} & 0.272 & 0.728 & 70.4 & $68^3$ & $2.6\times10^6$ & 0.05 & 4587 & 2000 & 10 & 1990\\
        \hline
    \end{tabular}
    \label{tab:simulations}
\end{table*}

\begin{figure*}
    \centering
    \includegraphics[width=\linewidth]{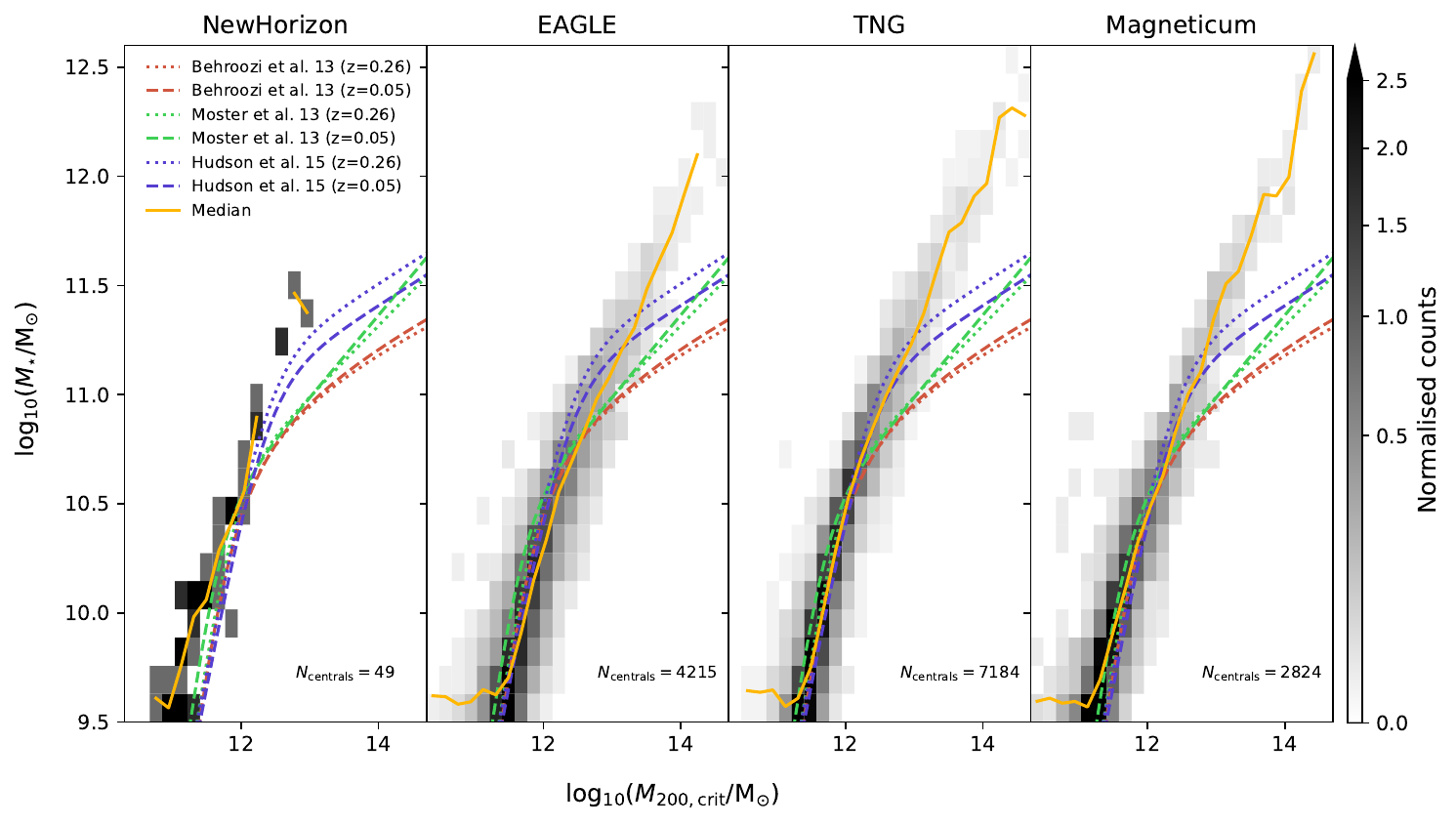}
    \caption{The galaxy stellar mass-halo mass relation for all central galaxies within our stellar mass range for each of our simulations, the sample sizes are given in the bottom right corner. The relation is plotted using 25 equally spaced bins for $9.5\leq\log_{10}(M_\star/$M$_\odot)\leq12.6$ and $10.3\leq\log_{10}(M_{\rm 200,\text{ crit}}/$M$_\odot)\leq14.7$. The dashed and dotted lines show observationally-derived relations from the literature for redshifts comparable to the simulation snapshots, the dotted lines correspond to $z=0.26$ (\textsc{NewHorizon} snapshot redshift) and the dashed lines correspond to $z=0.05$ (\textsc{EAGLE}, \textsc{TNG} and \textsc{Magneticum} snapshot redshifts). The median relation for central galaxies in each simulation is illustrated with an amber line. On average we see that the relations for \textsc{EAGLE}, \textsc{TNG} and \textsc{Magneticum} are similar, however, \textsc{NewHorizon} is missing high mass halos and contains more stars for a given halo mass than the other simulations. Note for this Figure we present the \textsc{AdaptaHOP} (for \textsc{NewHorizon}) and \textsc{SubFind} (for \textsc{EAGLE}, \textsc{TNG} and \textsc{Magneticum}) based galaxy stellar masses ($M_{\scriptstyle\star}$) for the galaxies. The counts have been normalised such that the volume under the histogram integrates to 1.}
    \label{fig:halo_mass_stellar_mass}
\end{figure*}

\begin{figure*}
    \centering
    \includegraphics[width=\linewidth]{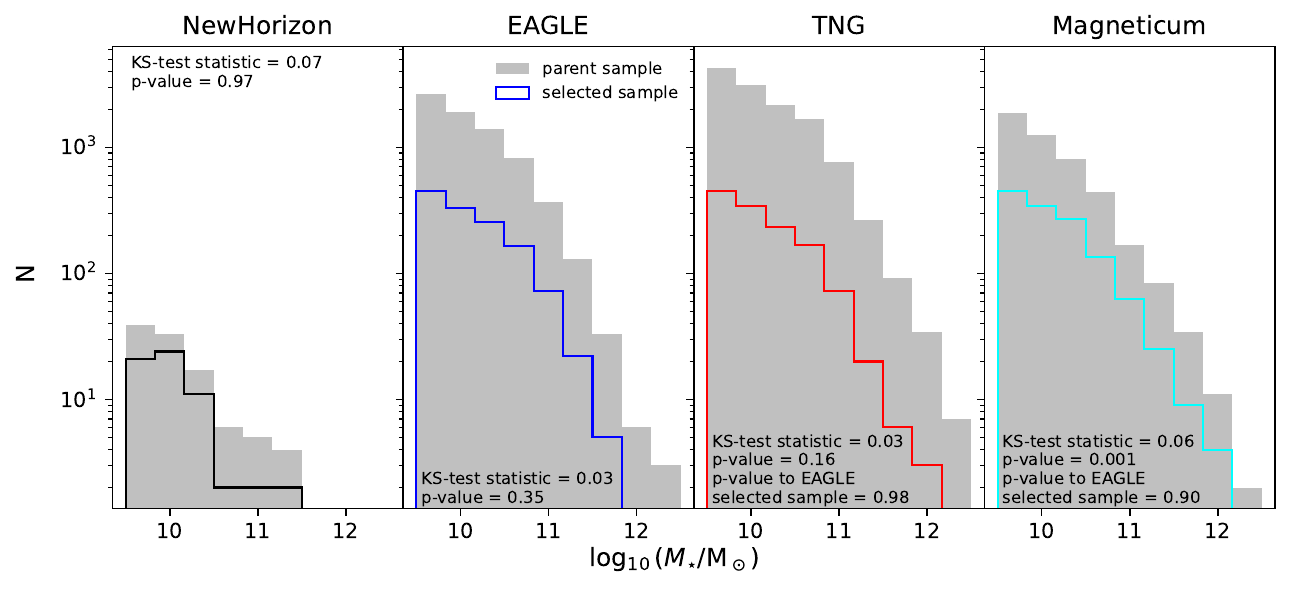}
    \caption{Stellar mass distributions of parent (grey) and selected samples (\textsc{NewHorizon} - black, \textsc{EAGLE} - blue, \textsc{TNG} - red and \textsc{Magneticum} - cyan) for each simulation. The $M_{\scriptstyle\star}\geq10^{9.5}$M$_{\scriptstyle \odot}$ range explored by our study is well sampled by the parent distribution for \textsc{TNG}, \textsc{EAGLE}, and \textsc{Magneticum}, but the smaller volume \textsc{NewHorizon} simulation has fewer galaxies at the high mass end. We include the results of the KS tests comparing the sample to the parent distributions. \textsc{EAGLE} roughly samples its parent distribution and \textsc{TNG} and \textsc{Magneticum} have been sampled to match \textsc{EAGLE}. The KS test results show that \textsc{TNG} remains similar to its parent, whereas \textsc{Magneticum} no longer resembles the parent distribution. Note for this Figure we present the \textsc{AdaptaHOP} (for \textsc{NewHorizon}) and \textsc{SubFind} (for \textsc{EAGLE}, \textsc{TNG} and \textsc{Magneticum}) based galaxy stellar masses ($M_{\scriptstyle\star}$) for the parent and subsampled galaxies.}
    \label{fig:sample_stellar_mass_distribution}
\end{figure*}

\begin{figure*}
    \centering
    \includegraphics[width=\linewidth]{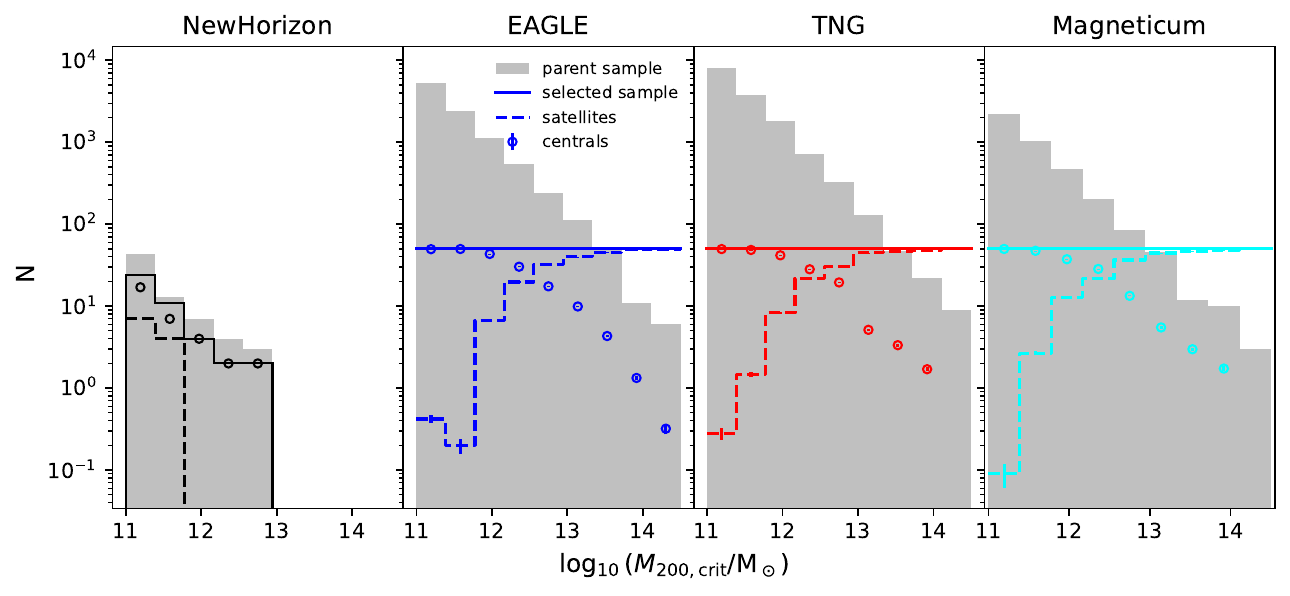}
    \caption{The solid lines show the total number of galaxies sampled in each of the halo mass bins for \textsc{NewHorizon} and the mean number sampled in each of the 100 Monte Carlo iterations for \textsc{EAGLE}, \textsc{TNG} and \textsc{Magneticum}. The solid lines illustrate that our Monte Carlo samples are flat in log halo mass by construction. We subsample 50 galaxies in each of 9 evenly spaced bins between $11\leq \log_{10}(M_{\scriptstyle200,\text{ crit}}/\text{M}_{\scriptstyle\odot})\leq14.5$.
    We also show the mean number of satellite galaxies per iteration using dashed lines and the mean number of centrals per iteration using circular points. We similarly depict the distributions of satellite and central galaxies for \textsc{NewHorizon}. The grey histograms show the parent distribution of log halo mass. The error bars give the standard error on the mean for each of the measurements obtained from the Monte Carlo iterations.}
    \label{fig:sample_halo_mass_distribution}
\end{figure*}

\subsection{Simulations}
\label{subsec:simulations}
To investigate the characteristics of tidal features in cosmological hydrodynamical simulations we explored four different simulations. These simulations evolve gas, stars and dark matter under gravity and hydrodynamics along with a range of models for subgrid processes to simulate the realistic hierarchical structure formation of galaxies.

For our study we selected four simulations, \textsc{NewHorizon}, \textsc{EAGLE RefL0100N1504}, \textsc{IllustrisTNG L75n1820TNG}, \textsc{Magneticum Pathfinder Box4-uhr} (from here on, \textsc{NewHorizon}, \textsc{EAGLE}, \textsc{TNG} and \textsc{Magneticum}). These four simulations, summarised in Table \ref{tab:simulations}, allow us to probe tidal features across a range of different resolutions, galaxy environments and subgrid physics models. As we approach the advent of large observational surveys of galaxies within the sensitivity to observe tidal features, it is important to characterise what each of these simulations predicts concerning these features. \textsc{TNG} and \textsc{EAGLE} probe a similar volume and therefore the same range of environments (isolated galaxies, groups and clusters). The \textsc{Magneticum} simulation probes an intermediate-sized box about one-third the volume of the two larger simulations and has a slightly lower stellar mass resolution. \textsc{NewHorizon} probes a much smaller volume with a much higher stellar mass resolution, allowing for more detailed tidal structures to be resolved by probing a more limited sample of galaxies. 

In this section we introduce each of the simulations, describing their mass resolution, volume and how they were calibrated. We discuss how galaxy structure, stellar mass and halo mass are measured in \S\ref{subsubsec:structure_and_substructure} and the simulations' stellar mass-halo mass (SMHM) relations in \S\ref{subsubsec:stellar_mass_halo_mass_relation}. We focus on these elements of the simulations as they are the salient differences for this study, directly impacting the number of stellar particles available to form tidal features and the halo masses that are covered by the simulations. The simulation subgrid models for star formation, stellar feedback and black hole feedback, which could indirectly contribute to the factors above and therefore, the tidal feature occurrence, are described in Appendix \ref{sec:app_sims}.

The \textsc{NewHorizon} \citep{duboisIntroducingNEWHORIZONSimulation2021} simulation is a zoom-in simulation from the parent \textsc{Horizon-AGN} simulation \citep{duboisDancingDarkGalactic2014}, which adopts the WMAP-7 cosmology \citep[][]{komatsuSevenyearWilkinsonMicrowave2011}. \textsc{NewHorizon} employs a spherical volume of $16^3$ cMpc$^3$ with varying dark matter resolution. The initial 10 cMpc radius has $m_{\rm DM}=1.2\times10^6\text{ M}_{\scriptstyle \odot}$, this high-resolution patch is embedded in spheres of decreasing mass-resolution of $10^7$, $8\times10^7$ and $6\times10^8\text{ M}_{\scriptstyle \odot}$ corresponding to spheres of radius $10.6$, $11.7$, and $13.9$ cMpc. The remaining volume is resolved at $5\times10^9\text{ M}_{\scriptstyle \odot}$ \citep{martinPreparingLowSurface2022}. The stellar mass resolution is $m_{\scriptstyle \star}=1.3\times10^4\text{ M}_{\scriptstyle \odot}$. Each stellar particle represents a simple stellar population with a Chabrier initial mass function \citep[IMF;][]{chabrierGalacticStellarSubstellar2003} ranging from 0.1 - 150 M$_{\scriptstyle \odot}$.

\textsc{Horizon-AGN} is calibrated to match the black hole density and the scaling relations between black hole mass and galaxy properties (albeit at a lower spatial resolution to \textsc{NewHorizon}). The \textsc{NewHorizon} simulation also reproduces the observed galaxy surface brightness as a function of stellar mass  \citep{duboisIntroducingNEWHORIZONSimulation2021}, without requiring any calibration for surface brightness. 

Table \ref{tab:simulations} shows that \textsc{NewHorizon} has stellar particles $\sim$2 orders of magnitude less massive than the other three simulations allowing it to resolve tidal features from higher mass ratio mergers. However, \textsc{NewHorizon} only covers field and group environments and does not include galaxy clusters of significant mass \citep[][]{duboisIntroducingNEWHORIZONSimulation2021}.

The \textsc{EAGLE} (Evolution and Assembly of GaLaxies and their Environments) project is a large set of cosmological hydrodynamical simulations. \textsc{EAGLE} contains simulations of cubic volume $12^3$, $25^3$, $50^3$ and $100^3$ cMpc$^{3}$.  The data which will be used in this study is from the reference model (\textsc{RefL0100N1504}) which has a volume of $100^3\text{ cMpc})^3$. The reference model is the simulation run with standard parameters and physics as described in \citet{schayeEAGLEProjectSimulating2015} and \citet{crainEAGLESimulationsGalaxy2015}. The simulation uses the cosmological parameters advocated by the Planck 2013 results \citep[][]{planckcollaborationPlanck2013Results2014}. The dark matter particle mass is $m_{\rm DM}=9.70 \times 10^{6} \text{ M}_{\scriptstyle \odot}$ and the stellar particle mass is $m_{\scriptstyle\star}=1.81\times10^{6} \text{ M}_{\scriptstyle \odot}$. 

The simulation was calibrated to match the galaxy stellar mass function (GSMF) and the relationship between black hole mass and galaxy stellar mass at $z\sim0$ \citep[details in][]{schayeEAGLEProjectSimulating2015}.

\textsc{IllustrisTNG} is a suite of simulations, with the following box volumes: $51.7^3$, $110.7^3$ and $302.6^3\text{ cMpc}^3$. We use simulation \textsc{L75n1820TNG}, which models the physics of dark matter and baryons in a $110.7^3\text{ cMpc}^3$ box. This simulation adopts the $\Lambda$CDM model fit by the Planck 2015 results \citep[]{adePlanck2015Results2016}. The dark and baryonic particle resolutions for \textsc{TNG100-1} are: $m_{\rm DM}=7.5\times10^6\text{ M}_{\scriptstyle \odot}$,
$m_{\scriptstyle\star}=1.4\times10^6\text{ M}_{\scriptstyle \odot}$. 

\textsc{TNG} is calibrated to match the star formation rate density, the GSMF and the stellar-to-halo mass relation, the total gas mass contained within the virial radius of massive groups, the stellar mass – stellar size and the black hole – galaxy mass relations all at $z = 0$, in addition to the overall shape of the cosmic star formation rate density at $z\lesssim10$ \citep{pillepichFirstResultsIllustristng2018}.

\textsc{Magneticum Pathfinder} simulations are a suite of cosmological hydrodynamical simulations, ranging in box size from $25.6^3$ to 3818$^3$ cMpc$^3$. We use the \textsc{Box4-uhr} simulation, which models the physics of dark matter and baryons in a $68^3$ cMpc$^3$ box. All simulations are performed with an updated version of the SPH code \textsc{GADGET-3} \citep{springelSimulationsFormationEvolution2005}. We use the Box4-uhr run, which adopts a WMAP-7 fit cosmology \citep{komatsuSevenyearWilkinsonMicrowave2011}. The mass resolution for stellar particles in this run is $m_{\scriptstyle \star}=2.6\times10^6$ M$_{\scriptstyle \odot}$ and $m_{\rm DM}=3.6\times10^7$ M$_{\scriptstyle \odot}$ for dark matter particles. 

The \textsc{Magneticum} simulations are calibrated to match the intracluster gas content of galaxy clusters, rather than matching any characteristics with individual galaxies.

\subsubsection{Identifying structure and measuring stellar mass}
\label{subsubsec:structure_and_substructure}

To study the properties of galaxies and their tidal features the particles in the simulations must be grouped into central galaxies and their satellites. For \textsc{TNG}, \textsc{EAGLE} and \textsc{Magneticum}, the galaxy and halo structure is computed using the baryonic version of the \textsc{SubFind} code \citep[][also \citealt{springelGADGETCodeCollisionless2001}]{dolagSubstructuresHydrodynamicalCluster2009}. The versions of structure finder used across these simulations are similar but not identical. In contrast, \textsc{NewHorizon} utilises \textsc{AdaptaHOP} for finding halos and subhalos \citep{tweedBuildingMergerTrees2009}. It is important to note that the stellar masses shown in Figure \ref{fig:halo_mass_stellar_mass} and Figure \ref{fig:sample_stellar_mass_distribution} are computed directly from each simulation's native structure finders. They are computed by summing the stellar masses of the particles assigned to each subhalo.

In general, \textsc{SubFind} and \textsc{AdaptaHOP} differ in how they assign particles to halos and subhalos: \textsc{SubFind} identifies substructure based on its local overdensity and gravitational boundedness. In its initial step \textsc{SubFind} uses a \textsc{Friends of Friends} (\textsc{FOF}) algorithm on dark matter particles to distribute them into halos, the baryons are then assigned to the halo of the nearest dark matter particle (if any). Substructures within each \textsc{FOF} halo are identified by searching for local density peaks, considering all particle types and resolution elements. Subhalos are identified using contours of isodensity that pass through limiting saddle points in the density field. Particles/resolution elements not gravitationally bound to the subhalo to which they have been assigned are removed by applying an iterative unbinding procedure. Finally, all particles and resolution elements that are unassigned to a subhalo are assigned to the central subhalo (the largest subhalo within a halo). One notable limitation of the \textsc{SubFind} method is that any resolution element/particle lying beyond the limiting isodensity contour is ignored, even if they are gravitationally bound to a subhalo \citep[e.g.][]{muldrewAccuracySubhaloDetection2011,canasIntroducingNewRobust2019}.

\textsc{AdaptaHOP} identifies substructure based on topology and does not carry out any unbinding. Particles are assigned to groups defined by peaks in the density field that are linked to other groups by saddle points in the density field. The substructure is then assigned by repeating this process with an increasing density threshold, resulting in smaller groups within groups. \textsc{AdaptaHOP} defines halos as a group of groups that exceeds 160 times the mean dark matter density. Groups within each halo are regrouped such that each subhalo has a mass smaller than the host halo.

The differences between the algorithms are non-trivial and will lead to differences in the stellar masses of galaxies. \textsc{AdaptaHOP} tends to identify fewer structures than \textsc{FOF}, i.e. several \textsc{FOF} halos are included in one \textsc{AdaptaHOP} halo, these differences can in-turn influence how \textsc{SubFind} assigns subhalos compared to \textsc{AdaptaHOP} \citep[][]{tweedBuildingMergerTrees2009}. We only use galaxy stellar masses derived from the simulation's native structure finding algorithm in Figure \ref{fig:halo_mass_stellar_mass} and Figure \ref{fig:sample_stellar_mass_distribution}, where we compare our visually-classified sample's stellar mass distribution to the parent stellar mass distribution in the simulation. In light of the differences between the algorithms we state that while the stellar mass distributions of both parent and child samples should not be compared trivially across the simulations, we feel comfortable in using the distributions to check how well our classified samples probe the stellar mass distributions of the simulations above a stellar mass of $10^{9.5}$M$_{\scriptstyle \odot}$. We also use the subhalo centres of potential as defined by \textsc{SubFind} and \textsc{AdaptaHOP} for the centres of our LSST-like mock images. Additionally, we rely on the halo assignment when loading particles to produce mock images as we only use particles assigned to the same halo. While for high halo masses (\textsc{AdaptaHOP} halo mass $\geq10^{13}$M$_{\scriptstyle\odot}$), \textsc{FOF} tends to assign $\gtrsim2$ times as many halos to a given collection of particles as \textsc{AdaptaHOP} \citep[][]{tweedBuildingMergerTrees2009}. We do not probe such high halo masses with \textsc{NewHorizon} therefore the differences in halo assignment will not have as significant an impact.

From \S\ref{subsec:results_mstar} onwards, when comparing the occurrence of tidal features as a function of stellar mass, to limit systematic biases between \textsc{AdaptaHOP} and \textsc{SubFind} derived masses we use the 30 pkpc spherical aperture, $M_{\star,\text{ 30 pkpc}}$, to measure the stellar masses. As discussed in \citet{schayeEAGLEProjectSimulating2015}, the 30 pkpc spherical aperture is comparable to the 2D Petrosian aperture often used in observational studies, making it a good choice for our aim to analyse the data in a manner directly comparable to observations. The spherical aperture measurement of stellar masses requires only the stellar particles assigned to a subhalo/galaxy within a 30 pkpc sphere centred on the subhalo's centre of potential to be included for the galaxy's stellar mass. 

In \S\ref{subsec:results_m200}, for our analysis of tidal feature fractions as a function of halo mass, we rely on the halo assignment from each simulation's native structure finding algorithms to define main halos, assign galaxies to main halos and the consequent measurement of $M_{\scriptstyle200,\text{ crit}}$ for the main halo. $M_{\scriptstyle200,\text{ crit}}$ is the total mass, including dark matter, gas and stars within the radius where the average density is $200$ times the critical density of the universe. For \textsc{TNG}, \textsc{EAGLE} and \textsc{Magneticum} which rely on \textsc{SubFind}, the calculation of $M_{\scriptstyle200,\text{ crit}}$ only relies on \textsc{FOF} for the identification of halos, the calculation of their mass is independent of the halo assignment. In contrast, \textsc{NewHorizon}'s \textsc{ADAPTHOP} measured $M_{\scriptstyle200,\text{ crit}}$ exclusively uses the resolution elements assigned to a particular halo. While each subhalo is assigned to a \textsc{FOF} identified host halo in \textsc{SubFind}, \textsc{AdaptaHOP}'s use of different density thresholds to separate halo and subhalo structure and the application of minimum density of 160 times the mean dark matter density for a structure containing substructures to be identified as a group leads to some subhalos in \textsc{NewHorizon} not being assigned to any parent halos.

\subsubsection{Stellar mass-halo mass relation}
\label{subsubsec:stellar_mass_halo_mass_relation}
The number of stars involved in interactions between halos determines the number of stars available to form tidal features. Therefore, it is important to consider the efficiency of star formation at a given halo mass for each of the simulations.

In Figure \ref{fig:halo_mass_stellar_mass} we show the SMHM distributions for each simulation, displaying all central galaxies within our stellar mass cut (sample sizes given by $N_{\rm centrals}$ in Figure \ref{fig:halo_mass_stellar_mass}). For comparison, we provide the observationally-constrained empirical SMHM relations from \citet[][constrained by stellar mass function, cosmic star formation rate and specific star formation rate]{behrooziAverageStarFormation2013} and \citet[][constrained by the stellar mass function]{mosterGalacticStarFormation2013} and the weak lensing derived relation from \citet{hudsonCFHTLenSCoevolutionGalaxies2015} and give the median SMHM relation for each of the simulations. We note that the median relation for each simulation appears to flatten for $M_{\scriptstyle\mathrm{200,\text{ crit}}}\lesssim10^{11}$ M$_{\scriptstyle\odot}$. This is driven by our lower stellar mass cut meaning that in this regime we only sample high mass halos due to not including the scatter to lower stellar masses \citep[e.g.][]{tekluMorphologyDensityRelation2017}.

\textsc{NewHorizon} qualitatively reproduces the shape of the observational relations for the halo mass range of the simulation. Relative to the empirical relations \textsc{NewHorizon} does deviate to higher stellar masses below the knee of the SMHM relation. This is at least partly due to the smaller simulation volume \citep{martinPreparingLowSurface2022}, resulting in under representation of group and cluster environments where star formation tends to be less efficient \citep[e.g.][]{garrison-kimmelStarFormationHistories2019,samuelExtinguishingFIREEnvironmental2022}. Another contributing factor to the higher star formation efficiency of \textsc{NewHorizon} could be the subgrid physics implemented (described in Appendix \ref{sec:app_sims}). We also note recent empirical studies of the low-mass end of the SMHM relation, constrained by the star formation rates of local group dwarf galaxies, have shown significant scatter to higher stellar masses for $M_{\scriptstyle\mathrm{200,\text{ crit}}}\lesssim 10^{10}$ M$_{\scriptstyle\odot}$, which is qualitatively similar to what is observed for \textsc{NewHorizon} \citep{olearyPredictionsStellartohaloMass2023}. 

\textsc{EAGLE}, \textsc{TNG} and \textsc{Magneticum} fall within the scatter of the observationally-constrained relations for $10^{11}$ M$_{\scriptstyle\odot}\lesssim M_{\scriptstyle\mathrm{200,\text{ crit}}}\lesssim10^{12}$ M$_{\scriptstyle\odot}$. For $M_{\rm\scriptstyle 200,\text{ crit}}\gtrsim10^{13.5}$ M$_\odot$ for \textsc{EAGLE} and $M_{\rm\scriptstyle 200,\text{ crit}}\gtrsim10^{12.5}$ M$_\odot$ for \textsc{TNG} and \textsc{Magneticum} we see that the centrals are sitting at systematically higher stellar masses than the observational relations. A factor contributing to this difference could be the differences between stellar mass measurements in simulations and observations. For example, \citet{remusAccretedNotAccreted2022} found better agreement between the cosmological simulations \textsc{EAGLE}, \textsc{Magneticum} and \textsc{TNG-300} and observational measurements of the SMHM relation \citep{kravtsovStellarMassHalo2018} when the light from the central galaxy and the intracluster light (ICL) around it was included in the observations. This deviation between simulations, observations, and models has been known for a while. \citet{kravtsovStellarMassHalo2018} showed that the observed SMHM relation deviates from the model predictions if the brightest cluster galaxy is measured out to large radii and their ICL is included. This leads in those cases to larger observed stellar masses at a given halo mass than predicted by \citet{behrooziAverageStarFormation2013} and \citet{mosterGalacticStarFormation2013}. This is found to be true for many simulations, where the SMHM relation from models and observations that exclude the ICL is usually reproduced if aperture cuts are applied to the simulations, but results in larger stellar masses if the stellar masses from the halo finders are used, that is the ICL and residuals from satellite galaxies are included in the stellar mass measurement of the brightest cluster galaxy (see for example \citet{remusAccretedNotAccreted2022} for \textsc{Magneticum}, \citet{pillepichFirstResultsIllustristng2018} for \textsc{IllustrisTNG}, \citet{contreras-santosCharacterisingIntraclusterLight2024} for The \textsc{Three Hundred Simulations}, and \citet{schayeFLAMINGOProjectCosmological2023a} for \textsc{FLAMINGOS}). And while the chosen apertures of 30-50 pkpc are somewhat artificial, it was recently demonstrated by \citet{broughPreparingLowSurface2024} by applying observational methods to separate brightest cluster galaxies from their ICL to mock images of simulations from Magneticum, \textsc{EAGLE}, \textsc{IllustrisTNG} and \textsc{Horizon-AGN}, that the best agreement is found for apertures of 50-100 pkpc, providing observational support for this previously artificial aperture cut.

There are small offsets between the SMHM relation for \textsc{EAGLE}, \textsc{TNG} and \textsc{Magneticum}, albeit smaller than the offset with \textsc{NewHorizon}. \textsc{EAGLE} is shifted to lower stellar masses compared to the other simulations and \textsc{Magneticum} shifted to marginally higher stellar masses than \textsc{TNG} at the low and high end of the relation. A detailed comparison between the models was carried out by \citet{remusAccretedNotAccreted2022}. They suggest that the likely cause of this is the star formation efficiencies of low-mass halos, where \textsc{Magneticum} is observed to convert more gas into galaxies. As a result of its SMHM relation, we expect NewHorizon to bring in more stars into mergers than the other three simulations for mergers containing halos of a given size.

\subsection{Sample selection}
\label{subsec:sample_selection}

We select all simulated galaxies with $M_{\star,\text{ 30 pkpc}}\geq10^{9.5}$M$_{\scriptstyle \odot}$. We make no distinction between centrals and satellites in our sampling, allowing the underlying distributions of centrals/satellites to carry over to our samples. At a stellar mass of $10^{9.5}$ M$_{\scriptstyle \odot}$ \textsc{Magneticum} resolves these galaxies with $\sim$ 1200 stellar particles, \textsc{EAGLE} with $\sim$ 1750, \textsc{TNG} with $\sim$ 2250 and \textsc{NewHorizon} with $\sim$ 24300. With any lower stellar mass threshold, there are fewer particles representing galaxies and our mock images will struggle to resolve significant tidal distortions to the stellar components of the galaxies. We also have few observational results to compare to below this stellar mass. 

We sampled our galaxies from similar redshift snapshots for \textsc{TNG}, \textsc{EAGLE} and \textsc{Magneticum}. However, for \textsc{NewHorizon} we used the two most recent available snapshots, taking care not to sample galaxies twice (Table \ref{tab:simulations}). The number of galaxies with $M_{\scriptstyle\star, \text{ 30 pkpc}}\geq10^{9.5}$ M$_{\scriptstyle \odot}$ in each simulation are given in Table \ref{tab:simulations}.

From the total sample of galaxies with $M_{\scriptstyle\star, \text{ 30 pkpc}}\geq10^{9.5}$ M$_{\scriptstyle \odot}$, we randomly subsample $2000$ galaxies to visually classify from each of \textsc{Magneticum} and \textsc{EAGLE} and 1907 galaxies from \textsc{TNG}. This is done so that the visual classification can occur over a reasonable timeframe. Within these galaxies, we note the presence of several galaxies which seemed to be very compact, point-like and have a stellar mass far higher than one would expect for such a compact object (see bottom right of Figure \ref{fig:mock_images_tidal_features}). We discarded these compact, point-like objects from our sample, the number discarded is given under $N_{\text{out}}$ in Table \ref{tab:simulations}. Our final sample sizes are 62 for \textsc{NewHorizon}, 1978 for \textsc{EAGLE}, 1826 for \textsc{TNG} and 1990 for \textsc{Magneticum}.

To study how tidal features depend on stellar mass in each of the simulations it is necessary to match the stellar mass distributions between the classified simulation samples to avoid a mass-biased comparison. We subsample from the classified \textsc{TNG} and \textsc{Magneticum} samples to match the stellar mass distribution of the parent \textsc{EAGLE} sample, which was calibrated to match the local GSMF \citep{furlongEvolutionGalaxyStellar2015}. Matching to \textsc{EAGLE}'s stellar mass distribution therefore makes our results more directly comparable with observational samples of tidal features. However, for \textsc{NewHorizon}, we do not subsample, leaving the sample as it is as we do not wish to reduce the sample size any further. 

To construct our stellar mass-matched sample, we fit the probability densities for the simulation parent stellar mass distributions using Gaussian kernel density estimation (KDE). Due to the unavailability of $M_{\scriptstyle\star, \text{ 30 pkpc}}$ measurements for the parent samples of \textsc{TNG} and \textsc{Magneticum} we use \textsc{SubFind}-derived $M_{\scriptstyle\star}$ values to fit the KDE for the parent stellar mass distributions from \textsc{EAGLE}, \textsc{TNG} and \textsc{Magneticum}. Using these KDEs we infer the likelihoods of sampling galaxies with particular $M_{\scriptstyle\star, \text{ 30 pkpc}}$ from each of the simulations. We then use these likelihoods to construct weights with which we sample the classified \textsc{TNG} and \textsc{Magneticum} galaxies to match the parent distribution of \textsc{EAGLE}. Our samples are constructed using the following steps:

\begin{enumerate}
    \item We draw, without replacement, N galaxies from \textsc{EAGLE}.
    \item We draw, without replacement, a sample of N galaxies from \textsc{TNG} and \textsc{Magneticum}, weighting the samples so that the likelihood of a galaxy being drawn is the same as it would be if we were sampling from the \textsc{EAGLE} parent stellar mass distribution.
    \item We redraw N galaxies from \textsc{TNG} and \textsc{Magneticum} until the distributions reasonably match that of \textsc{EAGLE}, by requiring that a KS test \citep{hodgesSignificanceProbabilitySmirnov1958} yield a p-value $>0.9$.
    \item The largest sample that could be subsampled from \textsc{TNG} and \textsc{Magneticum}, while satisfying condition (iii) is N=1300.
\end{enumerate}

The exact weights we used for subsampling from our classified \textsc{TNG} and \textsc{Magneticum} galaxies are as follows:

\begin{flalign}
W_{\scriptstyle\textsc{TNG}}(M_{\scriptstyle\star, \text{ 30 pkpc}}) &= P_{\scriptstyle\textsc{EAG.}}(M_{\scriptstyle\star, \text{ 30 pkpc}})/P_{\scriptstyle\textsc{TNG} }(M_{\scriptstyle\star, \text{ 30 pkpc}})&\\
W_{\scriptstyle\textsc{Mag.}}(M_{\scriptstyle\star, \text{ 30 pkpc}}) &= P_{\scriptstyle\textsc{EAG.}}(M_{\scriptstyle\star, \text{ 30 pkpc}})/P_{\scriptstyle\textsc{Mag.} }(M_{\scriptstyle\star, \text{ 30 pkpc}})&
\end{flalign}

Where $W_{\scriptstyle\textsc{TNG.}}(M_{\scriptstyle\star, \text{ 30 pkpc}})$ and $W_{\scriptstyle\textsc{Mag.}}(M_{\scriptstyle\star, \text{ 30 pkpc}})$, are the weights assigned to a \textsc{TNG} and \textsc{Magneticum} galaxy respectively, based on their $M_{\scriptstyle\star, \text{ 30 pkpc}}$. $P_{\scriptstyle\textsc{EAG.}}(M_{\scriptstyle\star, \text{ 30 pkpc}})$, $P_{\scriptstyle\textsc{TNG.} }(M_{\scriptstyle\star, \text{ 30 pkpc}})$ and $P_{\scriptstyle\textsc{Mag.} }(M_{\scriptstyle\star, \text{ 30 pkpc}})$ are the probability density functions from which we can calculate the likelihoods of galaxies with a particular $M_{\scriptstyle\star, \text{ 30 pkpc}}$ being sampled from the parent distribution of the respective simulations.

Figure \ref{fig:sample_stellar_mass_distribution} shows the stellar mass distributions of the parent samples for each of the simulations in grey histograms and the distributions of our selected sample in coloured histograms, black for \textsc{NewHorizon}, blue for \textsc{EAGLE}, red for \textsc{TNG} and cyan for \textsc{Magneticum}. \textsc{TNG} and \textsc{Magneticum} are mass-matched to \textsc{EAGLE}. The KS-test p-values for the matching of the $M_{\scriptstyle\star, \text{ 30 pkpc}}$ are 0.98 for \textsc{EAGLE} and \textsc{TNG} and 0.90 for \textsc{EAGLE} and \textsc{Magneticum}. 

We also compare our classified samples with the parent distributions from which they were sampled. This is done by using a KS-test to compare their distributions in structure finder-defined stellar mass ($M_{\scriptstyle\star}$). The high p-value of 0.35 for our \textsc{EAGLE} sample is expected as we have not attempted to adjust the sampling with respect to its parent distribution, we see \textsc{TNG} (p=0.16) and \textsc{Magneticum} (p=0.001) are less like their parent distributions as they have been sampled to match the 1300 galaxies sampled from \textsc{EAGLE}. We note that the conclusions we draw are not qualitatively changed by stellar mass-matching the samples.

For our halo mass-matched comparison we construct a flat distribution of log halo masses, from $11$ $\leq \log_{10}(M_{\scriptstyle200,\text{ crit}}/$M$_{\scriptstyle\odot})\leq14.5$ for \textsc{EAGLE}, \textsc{TNG} and \textsc{Magneticum}. Due to the small sample size of \textsc{NewHorizon} we do not attempt to perform any subsampling for the simulation. There is greater degeneracy amongst the parent halo masses of the galaxies as many galaxies can belong to the same halo. This makes it more difficult to match the populations with the weighted sampling method described above, so we opt for a Monte Carlo-based approach. To perform this we subsample our classified sample 100 times. We choose a linear bin spacing in $\log_{10}(M_{\scriptstyle200,\text{ crit}}/$M$_{\scriptstyle \odot})$, such that our tidal feature fraction with halo mass measurement would have a sufficient number of galaxies in each bin whilst having a large enough number of bins to resolve trends with halo mass. We found that a subsample of 450 galaxies per simulation, divided across 9 linearly spaced bins in log halo mass was appropriate for our purposes.

The leftmost panel in Figure \ref{fig:sample_halo_mass_distribution} shows our \textsc{NewHorizon} sample with respect to halo mass with a solid line, we note that \textsc{NewHorizon} does not include higher mass halos, equivalent to galaxy group and cluster mass systems, due to the small volume of this simulation. Furthermore, four of the galaxies in our \textsc{NewHorizon} sample had no parent halo assigned, reducing our sample to 58 galaxies for this part of the analysis. We also consider the number of central and satellite galaxies in our haloes. The central galaxy is the subhalo consisting of the largest number of particles within a parent halo by the structure finder, the other galaxies in the halo are defined to be satellites. In Figure \ref{fig:sample_halo_mass_distribution} the dashed line shows the number of satellite galaxies in each halo mass bin and the open circles show the number of central galaxies. The three rightmost panels in Figure \ref{fig:sample_halo_mass_distribution} illustrate the flat sampling of galaxies with respect to their parent halo mass for \textsc{EAGLE}, \textsc{TNG} and \textsc{Magneticum} that we apply where analysing the halo mass and central and satellite galaxy properties of our sample. The solid lines in each of these panels show the 50 galaxies sampled in each of the halo mass bins. The mean number of satellites in each bin is shown using a dashed line and the mean number of centrals is shown using open circles. We see that the number of satellites in each bin increases with increasing halo mass whereas the number of centrals decreases with halo mass. This is expected given that field galaxies are classified as centrals while group and clusters contain many satellite galaxies orbiting around one central galaxy.

\subsection{Mock images}
\label{subsec:mock_images}

Following the methods of \citet{martinPreparingLowSurface2022} we produce mock images that match the $0.2''$/pixel spatial resolution and expected 10-year surface brightness limits of LSST: $\mu_g\sim30.3$ mag arcsec$^{-2}$, $\mu_r\sim30.3$ mag arcsec$^{-2}$ and $\mu_i\sim29.7$ mag arcsec$^{-2}$(P. Yoachim; private communication, 3$\sigma$, $10''\times10''$). We produce colour images using the methods of \citet{luptonPreparingRedGreenBlueImages2004}. The mock images in our study are placed at $z=0.025$. The distance this corresponds to varies with the simulations adopted cosmology but is on average $\sim105$ pMpc. The mock \textit{g}, \textit{r} and \textit{i} images are made to be squares of 248 pkpc $\times$ 248 pkpc $\simeq480''\times480''$ $=$ 2400 pixels $\times$ 2400 pixels.

Examples of mock-observed images are shown in Figure \ref{fig:mock_images_tidal_features}. Described briefly the construction of mock images involves:
\begin{enumerate}
    \item Extracting all the stellar particles associated with a group (\textsc{FOF} or \textsc{AdaptaHOP} identified halo, depending on the simulation)  within a 1 Mpc cube of the simulated galaxy and calculating spectral energy distributions (SED) for each stellar particle using simple stellar population models \citep{bruzualStellarPopulationSynthesis2003} extrapolated to the age and metallicity of each stellar particle. Dust attenuation is accounted for in \citet{martinPreparingLowSurface2022}'s method but is not included in the mock images created here. The brightness for each of the stellar particles is determined by the redshift of the galaxy and the convolution between the SED and the transmission function for the relevant LSST filters \citep{olivierOpticalDesignLSST2008a}. Using this method we produce apparent magnitudes for the LSST $g$, $r$ and $i$ filters.
    \item Smoothing the particle distribution in regions where there is a potential for undersampling in the final image. This smoothing is necessary to remove unrealistic variations in flux between pixels in the mock image. This is done by subdividing stellar particles into smaller mass particles with positions distributed normally around the original particle, with a standard deviation equal to the distance to the 5th nearest neighbour of the stellar particle being smoothed \citep[][see also \citealt{merrittMissingOutskirtsProblem2020}]{martinPreparingLowSurface2022}. 
    \item Produce mock images by collapsing the cube along one of its axes with a grid size of $0.2''\times0.2''$ \citep[the LSST pixel size;][]{ivezicLSSTScienceDrivers2019a}. The apparent magnitudes of the light from each of the pixels are measured accounting for the dimming due to the redshift ($z=0.025$). 
    \item To model the observational effects of seeing, the image is then convolved with the \textit{g}-band Hyper-Suprime Cam point spread function \citep{montesBuildupIntraclusterLight2021a} rescaled to match the LSST pixel size.
    \item To match observational background noise, Gaussian noise is added using the empirical relation between the standard deviation of the background noise and the limiting surface brightness, given in \citet[][]{romanGalacticCirriDeep2020}.
    \begin{equation}
        \sigma_{\rm{noise}}=\frac{10^{-0.4\mu_{\rm band}^{\rm lim}(n\sigma, \Omega\times\Omega)}\text{pix}\Omega}{n}
    \end{equation}
    Where $\Omega$ is the size of one side of the box over which the surface brightness limit is computed, $=10''$, pix is the pixel scale in arcseconds per pixel, $=0.2$ arcseconds/pixel, $n$ is the number of Gaussian standard deviations for the surface brightness limits, $=3$ and $\mu_{\rm band}^{\rm lim}$ is the limiting surface brightness for a particular photometric band. This model for noise intends to include all possible spatially invariant sources of noise, including instrumental noise, unresolved background sources, scattered light etc. without explicitly modelling them. However, we note that this model does not include spatially varying sources of noise, e.g. sky gradients, diffraction spikes, imperfect sky subtraction or galactic cirrus.
\end{enumerate}

Martin et al. (in prep) have shown using Sersic fits to mock images and galaxies in the HSC-SSP observed Cosmos field that this technique for producing mock images does not induce any measurable biases in galaxy structure measured from the mock images. They found that the variance between simulated and observed galaxies is mostly driven by the differences between the galaxy evolution models.

\subsection{Tidal feature classification}
\label{subsec:tidal_feature_classification}

We carry out our classification using single \textit{g}, \textit{r} and \textit{i} band images along with the combined colour image, to give ourselves the best chance of detecting and correctly classifying these faint features. The images were viewed at a fixed range of contrast and brightness as shown in Figure \ref{fig:mock_images_tidal_features}. The classification is carried out by the lead author. To classify the tidal features we chose to follow a classification scheme similar to that used in \citet{bilekCensusClassificationLowsurfacebrightness2020} and \citet{desmonsGalaxyMassAssembly2023}. The categories include:
\begin{itemize}
    \item \textbf{Streams/Tails}: Prominent, elongated structures orbiting or expelled from the host galaxy. These usually have similar colours to the host galaxy
    \item \textbf{Shells}: Concentric radial arcs or ring-like structures around a galaxy.
    \item \textbf{Plumes or Asymmetric Stellar Halos}: Diffuse features in the outskirts of the host galaxy, lacking well-defined structures like stellar streams or tails.
    \item \textbf{Double nuclei}: Two clearly separated galaxies within the mock image where merging is evident through the presence of tidal features.
\end{itemize}

In this scheme galaxies are allowed to host more than one feature type. A feature is allocated to the galaxy to which it appears to connect to, e.g. a tail/stream is allocated to the galaxy from which it stems, a shell is associated with the galaxy on which it is centred and a asymmetric halo is associated with the galaxy possessing the halo. In cases where features can be associated with multiple objects, e.g. a tail/stream between two galaxies (a tidal bridge) or an asymmetric halo enveloping a double nucleus, the tidal feature is counted for both objects. Following \citet{desmonsGalaxyMassAssembly2023} we have merged the categories of streams and tails as they are easily mistaken for one another at the surface brightness depth of LSST \citep{martinPreparingLowSurface2022}. We also use a system of confidence levels in our classification detailed in Table \ref{tab:confidence_levels}. This system is similar to that followed by \citet{atkinsonFAINTTIDALFEATURES2013}, but we group their confidence levels 3 and 4 which are represented as confidence level 3 here. Example features are shown in Figure \ref{fig:mock_images_tidal_features}.

\begin{table}
    \centering
    \begin{tabular}{lp{4cm}|l}
    \hline\\
        Confidence Level ($conf.$) & Description\\
    \hline
       0  & No tidal feature detected.\\
    \hline
       1  & Hint of tidal feature detected, classification difficult.\\
    \hline
       2  & Even chance of correct classification of tidal feature presence and/or morphology.\\
    \hline
       3  & High likelihood of the tidal feature being present and morphology being obvious.\\
    \hline
    \end{tabular}
    \caption{Descriptions of the classification confidence levels used for our analysis.}
    \label{tab:confidence_levels}
\end{table}

\subsection{Calculating the errors on our counts and fractions}
\label{subsec:caculating_errors}

We estimate the binomial errors on the tidal feature fractions presented here using $1\sigma\simeq0.683$ confidence levels computed using a Bayesian beta distribution generator for binomial confidence intervals described in detail  in \citet{cameronEstimationConfidenceIntervals2011}. This approach is robust against small and intermediate sample sizes, which is particularly important due to the small size of the \textsc{NewHorizon} sample.

\section{Results}
\label{sec:results}

\begin{figure*}
    \centering
    \includegraphics[width=\linewidth]{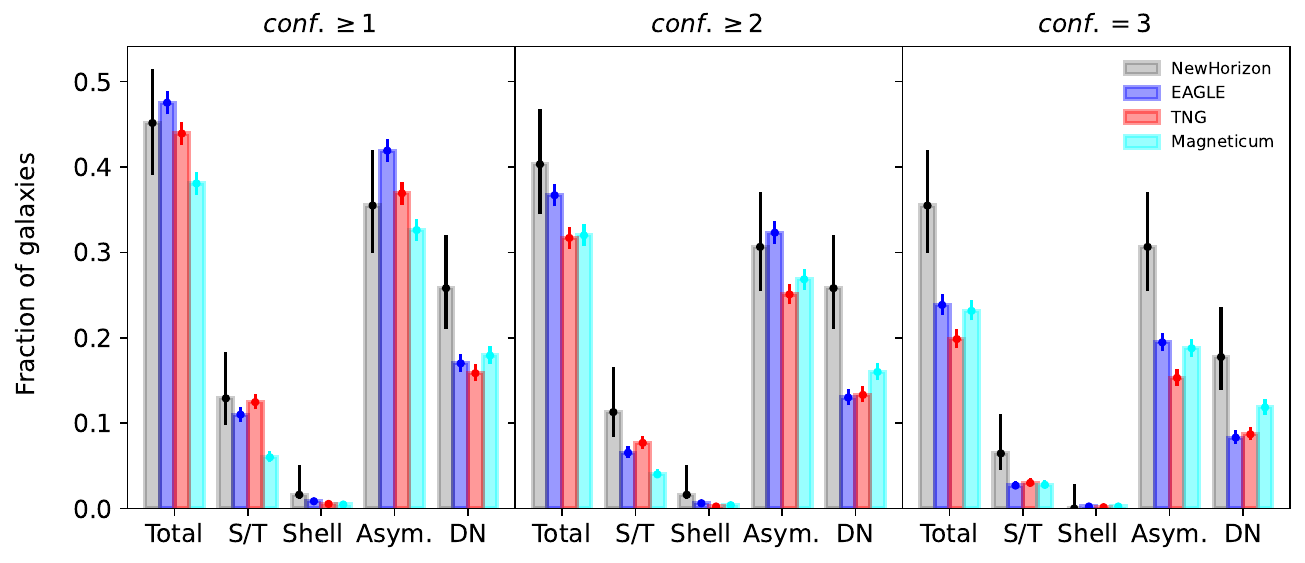}
    \caption{The frequencies of the tidal feature morphologies for each simulation: \textsc{NewHorizon} in grey, \textsc{EAGLE} in blue, \textsc{TNG} in red and \textsc{Magneticum} in cyan. From left to right the panels show the results for each confidence threshold: $conf.\geq1$, $conf.\geq2$ and $conf.=3$ (described in Table \ref{tab:confidence_levels}). The fractions are given for 5 feature categories: total tidal feature fraction (Total), stream/tail (S/T), shell, asymmetric halo (Asym.) and double nucleus (DN). The error bars give the $1\sigma$ binomial confidence levels. The occurrence of features relative to one another is similar across each confidence level, the fractions of galaxies exhibiting features drop as the confidence level increases. Generally, the simulations are in good agreement, with \textsc{NewHorizon}, tending to have a higher fraction tidal features that the other simulations.}
    \label{fig:tidal_feature_frequency}
\end{figure*}

\begin{table*}
\centering
\caption{The tidal feature fractions for every feature type and classification confidence level ($conf.\geq1$, $conf.\geq2$ and $conf.=3$, described in Table \ref{tab:confidence_levels}). The 5 feature categories presented are: total tidal features, stream/tail, shell, asymmetric halo and double nucleus. The rightmost column gives the overall mean fraction across the four simulations. The uncertainties give the $1\sigma$ binomial confidence intervals on each of the fractions. ${}^{*}$ \textsc{NewHorizon} is excluded from the calculation of the mean shell fraction as there is only one feature in that sample.}
\begin{tabular}{c|c|c|c|c|c|c}
\hline
Tidal features  & Confidence level & \textsc{NewHorizon} & \textsc{EAGLE} & \textsc{TNG} &  \textsc{Magneticum} & Overall mean\\
\hline
Total Tidal Features & $conf.\geq1$ & $0.45\pm0.06$ & $0.48\pm0.01$ & $0.44\pm0.01$ & $0.38\pm0.01$ & $0.44\pm0.07$\\
\\
 & $conf.\geq2$ & $0.40\pm0.06$ & $0.37\pm0.01$ & $0.32\pm0.01$ & $0.32\pm0.01$ & $0.35^{+0.07}_{-0.06}$\\
 \\
 & $conf.=3$ & $0.35\pm0.06$ & $0.24\pm0.01$ & $0.20\pm0.01$ & $0.23\pm0.01$ & $0.26^{+0.07}_{-0.06}$\\
 \hline
Stream/Tail & $conf.\geq1$ & $0.13^{+0.05}_{-0.03}$ & $0.110^{+0.009}_{-0.008}$ & $0.124^{+0.01}_{-0.009}$ & $0.060^{+0.007}_{-0.006}$ & $0.11^{+0.06}_{-0.03}$\\
\\
 & $conf.\geq2$ & $0.11^{+0.05}_{-0.03}$ & $0.065^{+0.008}_{-0.006}$ & $0.077^{+0.008}_{-0.007}$ & $0.040^{+0.006}_{-0.005}$ & $0.07^{+0.05}_{-0.03}$ \\
 \\
 & $conf.=3$ & $0.06^{+0.05}_{-0.02}$ & $0.027^{+0.005}_{-0.004}$ & $0.03^{+0.005}_{-0.004}$ & $0.028^{+0.005}_{-0.004}$ & $0.04^{+0.05}_{-0.02}$\\
 \hline
Shell & $conf.\geq1$ & $0.016^{+0.04}_{-0.005}$ & $0.008^{+0.003}_{-0.002}$ & $0.005^{+0.003}_{-0.001}$ & $0.005^{+0.003}_{-0.001}$ & $0.006^{+0.001}_{-0.0007}{}^{*}$ \\
\\
 & $conf.\geq2$ & $0.016^{+0.04}_{-0.005}$ & $0.006^{+0.003}_{-0.001}$ & $0.0023^{+0.002}_{-0.0007}$ & $0.004^{+0.003}_{-0.001}$ & $0.0041^{+0.002}_{-0.0005}{}^{*}$ \\
 \\
 & $conf.=3$ & $0.0^{+0.03}$ & $0.0023^{+0.002}_{-0.0007}$ & $0.0015^{+0.002}_{-0.0005}$ & $0.0023^{+0.002}_{-0.0007}$ & $0.0020^{+0.001}_{-0.0004}{}^{*}$ \\
\hline
Asymmetric Halo & $conf.\geq1$ & $0.35\pm0.06$ & $0.42\pm0.01$ & $0.37\pm0.01$ & $0.33\pm0.01$ & $0.37^{+0.07}_{-0.06}$  \\
\\
 & $conf.\geq2$ & $0.31^{+0.06}_{-0.05}$ & $0.32\pm0.01$ & $0.25\pm0.01$ & $0.27\pm0.01$ & $0.29^{+0.07}_{-0.06}$ \\
 \\
 & $conf.=3$ & $0.31^{+0.06}_{-0.05}$ & $0.19\pm0.01$ & $0.153^{+0.01}_{-0.009}$ & $0.19\pm0.01$ & $0.21^{+0.07}_{-0.05}$ \\
\hline
Double Nucleus & $conf.\geq1$ & $0.26^{+0.06}_{-0.05}$ & $0.17\pm0.01$ & $0.16\pm0.01$ & $0.18\pm0.01$ & $0.19^{+0.07}_{-0.05}$ \\
\\
 & $conf.\geq2$ & $0.26^{+0.06}_{-0.05}$ & $0.130^{+0.01}_{-0.009}$ & $0.133^{+0.01}_{-0.009}$ & $0.16\pm0.01$ & $0.17^{+0.07}_{-0.05}$ \\
 \\
 & $conf.=3$ & $0.18^{+0.06}_{-0.04}$ & $0.083^{+0.008}_{-0.007}$ & $0.087^{+0.008}_{-0.007}$ & $0.118^{+0.01}_{-0.008}$ & $0.12^{+0.06}_{-0.04}$\\
\hline
\label{tab:tidal_feature_fraction}
\end{tabular}
\end{table*}

We present the results of our classification in Figure \ref{fig:tidal_feature_frequency} and Table \ref{tab:tidal_feature_fraction}. Figure \ref{fig:tidal_feature_frequency} illustrates the fractions of galaxies exhibiting each kind of tidal feature for the four simulations: \textsc{EAGLE} in blue, \textsc{TNG} in red, \textsc{NewHorizon} in grey and \textsc{Magneticum} in cyan, for each of the confidence levels given in Table \ref{tab:confidence_levels}. The error bars show the $1\sigma$ binomial errors on the fractions. We find excellent agreement between the four simulations at each confidence level. The differences between the tidal feature fractions are much smaller than the change in the  fractions with feature morphology or with confidence level. The simulations all agree on the relative frequencies of tidal features, namely, asymmetries in the halo are the most common feature followed by double nuclei, streams/tails and lastly shells. 

As expected, the fraction of galaxies exhibiting tidal features decreases with an increasing classification confidence threshold ($conf.$). The agreement between \textsc{Magneticum}, \textsc{TNG} and \textsc{EAGLE} gets stronger with the increasing confidence threshold, indicating that as we look at tidal features that are obvious/well resolved in the 3 simulations we have an increasing agreement. 

We observe the inverse trend with \textsc{NewHorizon}, whilst still agreeing well with the frequencies of tidal features found in the other simulations. For $conf.\geq2$, \textsc{NewHorizon} systematically sits at higher fractions of tidal features than any of the other simulations. Table \ref{tab:tidal_feature_fraction} shows that between $conf.\geq1$ and $conf.=3$, \textsc{NewHorizon}'s tidal feature fraction drops by $0.1$, whereas in \textsc{EAGLE} and \textsc{TNG}, it drops by $0.24$ and in \textsc{Magneticum} the fraction drops by $0.15$. The smaller change in tidal feature fractions when moving from the lowest to the highest confidence level shows that tidal features detected in \textsc{NewHorizon} tend to be higher confidence than the other simulations.

The qualitative threshold of confidence levels does not change the relative frequencies of tidal features within simulations, they act as a trade-off between the detectability of a tidal feature by visual classification and the accuracy of its classification. For this reason, we tread a middle path between detectability and accuracy for the remainder of the results and present the moderate confidence results (i.e. $conf.\geq2$), noting that the choice of confidence level does not qualitatively change the conclusions we draw. 

At $conf.\geq2$ the total tidal feature fractions (given in Table \ref{tab:tidal_feature_fraction}) are: $0.40\pm0.06$ for \textsc{NewHorizon}, $0.37\pm0.01$ for \textsc{EAGLE}, $0.32\pm0.01$ for \textsc{TNG} and $0.32\pm0.01$ for \textsc{Magneticum}.

We study the tidal feature fraction as a function of stellar and halo mass. In \S\ref{subsec:results_mstar} we present the results of our visual classification in a mass-matched comparison of \textsc{EAGLE}, \textsc{TNG} and \textsc{Magneticum} and the entire sample of 62 galaxies from the higher resolution \textsc{NewHorizon} simulation. In \S\ref{subsec:results_m200} we study the trends with halo mass, considering this as a proxy for galaxy environment \citep[e.g.][]{yangGalaxyOccupationStatistics2005}.

\subsection{Tidal features and stellar mass}
\label{subsec:results_mstar}

\begin{table*}
    \centering
    \caption{The mean galaxy stellar masses for each of the tidal feature categories, from top to bottom: all tidal features, stream/tail, shell, asymmetric halo and double nucleus, for the four simulations, from left to right: \textsc{NewHorizon}, \textsc{EAGLE}, \textsc{TNG} and \textsc{Magneticum}. The final column shows the overall mean over the four simulations for the tidal feature host mass. The uncertainties show the standard deviations on the means. $^*$ \textsc{NewHorizon} is excluded from the calculation of the average shell host mass as there is only one feature in that sample.}
    \begin{tabular}{p{3cm}c|c|c|c|c|c}
    \hline
        Mean masses & \textsc{NewHorizon} & \textsc{EAGLE} & \textsc{TNG} & \textsc{Magneticum} & Average\\ 
        
        [$\log_{10}(M_{\scriptstyle\star, \text{ 30 pkpc}}/$M$_{\scriptstyle \odot})$] & & & \\
    \hline
        All tidal features & $10.3\pm0.8$ & $10.3\pm0.6$ & $10.4\pm0.7$ & $10.4\pm0.9$ & $10.3\pm0.4$\\
    \hline
        Stream/Tail & $10.9\pm0.4$ & $10.6\pm0.4$ & $10.7\pm0.5$ & $10.7\pm0.4$ & $10.7\pm0.2$\\
    \hline
        Shell & $10.4$ & $11.0\pm0.1$ & $11.1\pm0.2$ & $11.1\pm0.4$ & $11.1\pm0.1^*$\\
    \hline
        Asymmetric Halo & $10.5\pm0.6$ & $10.5\pm0.6$ & $10.6\pm0.6$ & $10.6\pm0.6$ & $10.6\pm0.3$\\
    \hline
        Double Nucleus & $10.6\pm0.6$ & $10.6\pm0.5$ & $10.7\pm0.6$ & $10.8\pm0.7$ & $10.7\pm0.3$\\
    \hline
    \end{tabular}
    \label{tab:tidal_feature_mass}
\end{table*}

\begin{figure*}
    \centering
    \includegraphics[width=\linewidth]{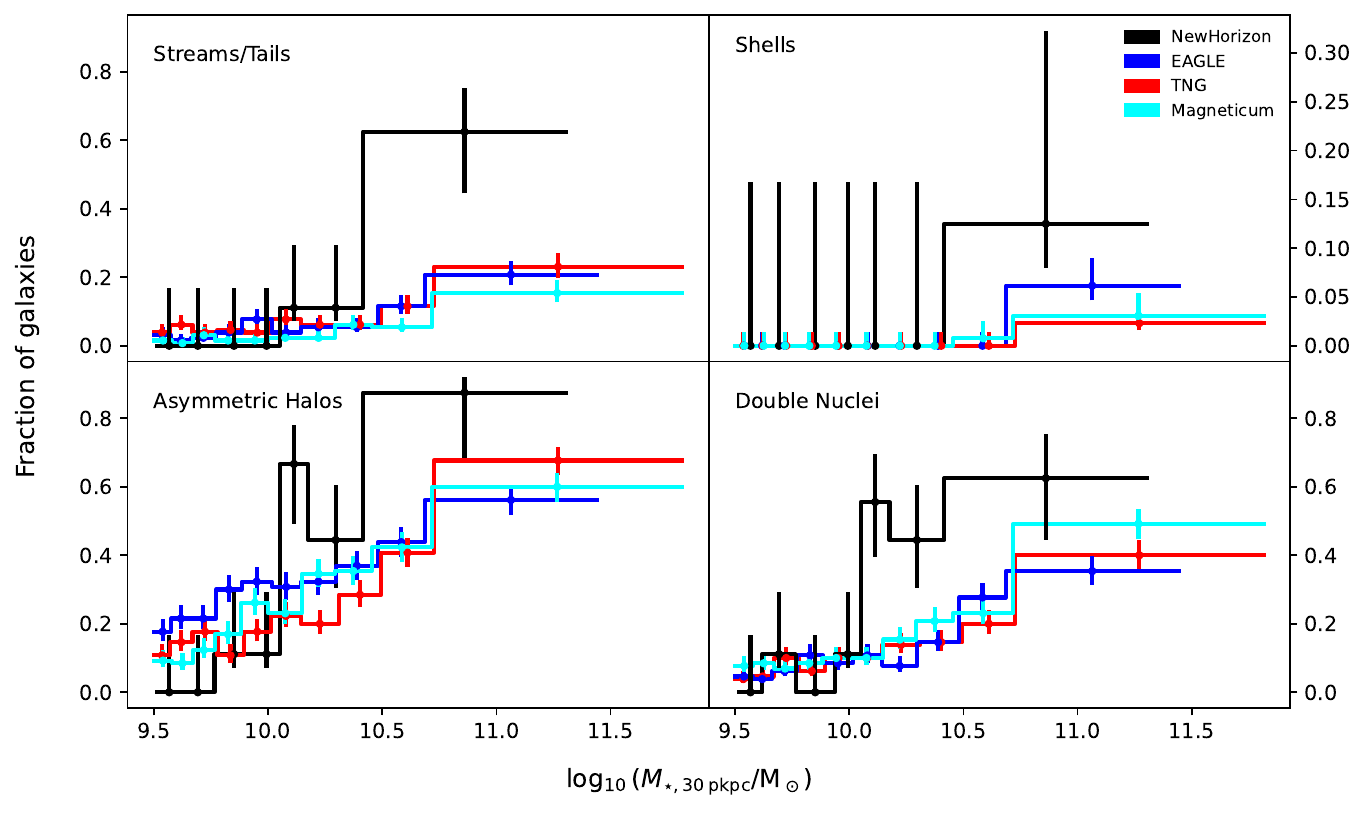}
    \caption{Distributions of the fraction of galaxies exhibiting at least one instance of a tidal feature over the range of stellar masses. From left to right, top to bottom we have the fractions of galaxies hosting: streams/tails, shells, asymmetric halos, and a double nucleus. The fractions for \textsc{NewHorizon} are shown in black, \textsc{EAGLE} in blue, \textsc{TNG} in red, and \textsc{Magneticum} in cyan. The error bars indicate the $1\sigma$ binomial errors. For each simulation, each mass bin contains a similar number of galaxies, $N_{\mathrm{bin},\textsc{TNG}} = N_{\mathrm{bin},\textsc{EAGLE}} = N_{\mathrm{bin}, \textsc{Magneticum}} = 130$ and $N_{\mathrm{bin}, \textsc{NewHorizon}}\simeq 9$. The distributions of tidal features as a function of stellar mass agree well across all simulations, and exceptionally well between the stellar mass-matched samples from \textsc{TNG}, \textsc{EAGLE} and \textsc{Magneticum}.}
    \label{fig:tidal_feature_fraction_mass}
\end{figure*}

Table \ref{tab:tidal_feature_mass} shows the average stellar masses of the tidal feature hosts for every simulation at the moderate classification confidence level ($conf.\geq2$). These are in good agreement across each of the simulations. The detected shells have the highest overall mean stellar masses across the simulations, $M_{\scriptstyle\star, \text{ 30 pkpc}}\sim10^{11.2}$ M$_{\scriptstyle \odot}$, not including \textsc{NewHorizon} which only has 1 shell in the classified sample. The other tidal features appear to have very similar mean stellar masses.

Figure \ref{fig:tidal_feature_fraction_mass} investigates the fraction of the galaxy population exhibiting tidal features above the LSST surface brightness limits as a function of stellar mass. We do this by constructing stellar mass bins with equal numbers of galaxies. For the stellar mass-matched samples of \textsc{EAGLE}, \textsc{Magneticum} and \textsc{TNG} we have 10 bins containing 130 galaxies each. For \textsc{NewHorizon} the 62 galaxies are distributed across 7 bins of $\sim9$ galaxies. We again see excellent agreement between the mass-matched simulations and broadly good agreement for \textsc{NewHorizon}, though we note that \textsc{NewHorizon} has systematically higher feature fractions, particularly for $M_{\scriptstyle \star}>10^{10}$ M$_{\scriptstyle \odot}$.

The top left panel of Figure \ref{fig:tidal_feature_fraction_mass} displays the stream/tail fraction - stellar mass distribution of galaxies for each simulation. All simulations exhibit a trend of increasing stream/tail fraction with increasing stellar mass, with very strong agreement between \textsc{TNG}, \textsc{EAGLE} and \textsc{Magneticum}. \textsc{NewHorizon} exhibits the same trend but has systematically higher fractions than the 3 other simulations. However, it is important to keep in mind the much smaller sample size (and therefore larger uncertainties) and higher resolution of \textsc{NewHorizon} when considering the significance of these offsets.

The top right panel shows the shell fraction - stellar mass distributions. There is a notable increase in the fraction of galaxies exhibiting shells at $M_{\star,\text{ 30 pkpc}}\gtrsim10^{11}$ M$_{\scriptstyle \odot}$ for \textsc{TNG}, \textsc{EAGLE} and \textsc{Magneticum}. The singular shell in \textsc{NewHorizon} does not provide sufficient information to determine any trends.

The asymmetric halo fraction - stellar mass distribution for each simulation is plotted in the bottom left panel of Figure \ref{fig:tidal_feature_fraction_mass}. Again we see excellent agreement between the simulations and a trend of increasing asymmetric halo fraction with increasing stellar mass. The rate of increase in tidal feature fraction with stellar mass is the largest here (from $\sim0.2$ to $\sim0.6$), indicating a strong dependence on stellar mass.  

Lastly, the bottom right panel of Figure \ref{fig:tidal_feature_fraction_mass} shows the double nucleus fraction for each of the stellar mass bins. We see a sharply increasing double nucleus fraction with increasing stellar mass for \textsc{TNG}, \textsc{EAGLE} and \textsc{Magneticum}, and excellent agreement between the three. For \textsc{NewHorizon}, there is a sharp increase with increasing stellar mass and an indication of a higher fraction in the highest mass bins.

\subsection{Tidal features and halo mass}
\label{subsec:results_m200}

\begin{figure*}
    \centering
    \includegraphics[width=\linewidth]{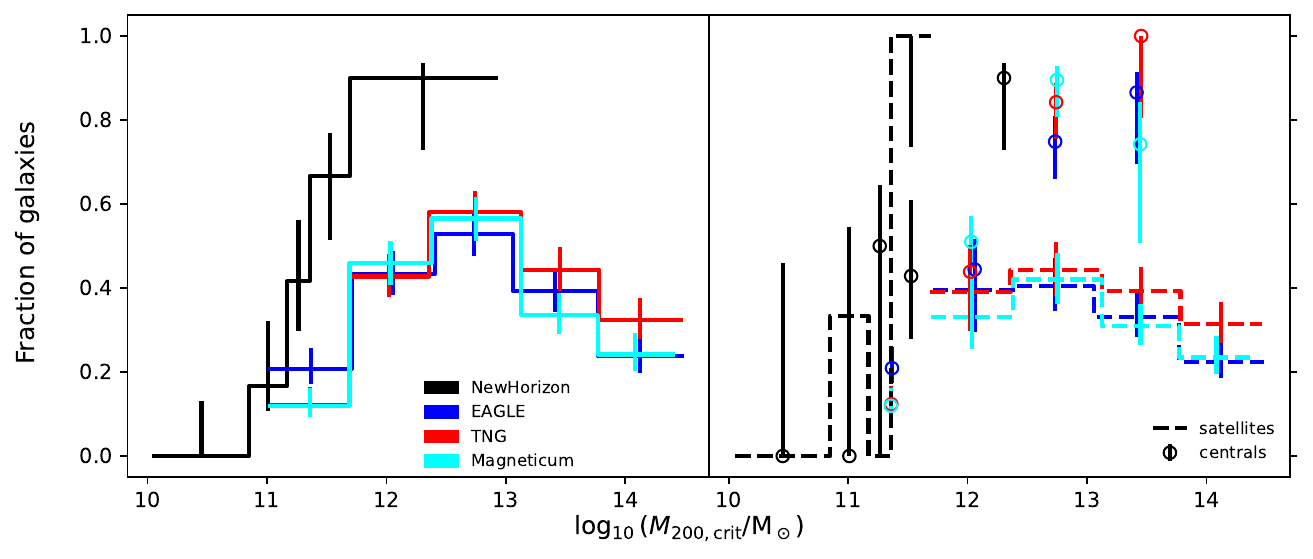}
    \caption{Distributions of the total tidal feature fractions as a function of halo mass. The left panel shows the total tidal feature fractions for each of our simulations and the right panel shows the total tidal feature fractions for the central and satellite galaxy populations. The results for \textsc{NewHorizon} are given in black, \textsc{EAGLE} in blue, \textsc{TNG} in red, and \textsc{Magneticum} in cyan. The \textsc{NewHorizon} data is given for the sample of 58 classified galaxies with assigned parent halos and each point represents the feature fraction for a halo mass bin containing $\simeq12$ galaxies. The points for \textsc{EAGLE}, \textsc{TNG} and \textsc{Magneticum} represent the mean of 100 Monte Carlo iterations, where each iteration subsampled 450 galaxies to the flat log halo mass distributions depicted in Figure \ref{fig:sample_halo_mass_distribution}. Each of the bins for \textsc{EAGLE}, \textsc{TNG} and \textsc{Magneticum} contain 90 galaxies, these galaxies are a varying mix of central and satellite galaxies as can be seen in Figure \ref{fig:sample_halo_mass_distribution}. The error bars show the mean $1\sigma$ binomial errors.}
    \label{fig:total_tidal_feature_fraction_halo_mass}
\end{figure*}

\begin{figure*}
    \centering\includegraphics[width=\linewidth]{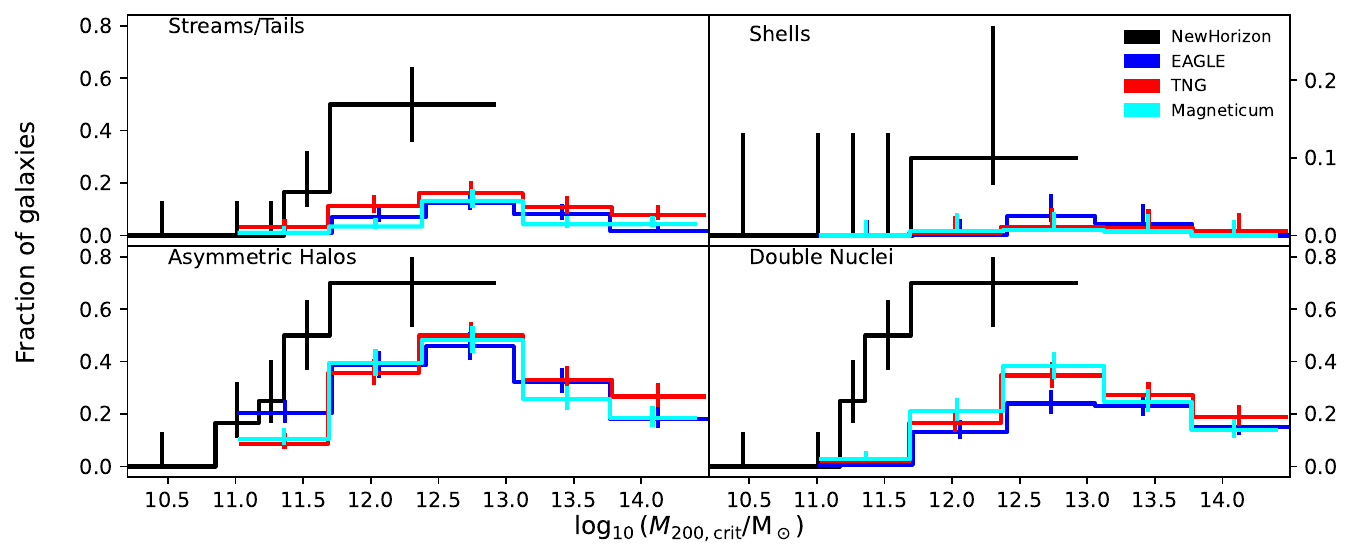}
    \caption{Distributions of tidal feature fractions as a function of halo mass. From left to right, top to bottom, we have the: stream/tail fraction, shell fraction, asymmetric halo fraction, and double nucleus fraction. The results for \textsc{NewHorizon} are given in black, \textsc{EAGLE} in blue, \textsc{TNG} in red, and \textsc{Magneticum} in cyan. The \textsc{NewHorizon} data is given for the sample of 58 classified galaxies with assigned parent halos and each point represents the feature fraction for a halo mass bin containing $\simeq12$ galaxies. The points for \textsc{EAGLE}, \textsc{TNG} and \textsc{Magneticum} represent the mean of 100 Monte Carlo iterations, where each iteration subsampled 450 galaxies to the flat log halo mass distributions depicted in Figure \ref{fig:sample_halo_mass_distribution}. Each of the bins for \textsc{EAGLE}, \textsc{TNG} and \textsc{Magneticum} contain 90 galaxies. The error bars show the mean $1\sigma$ binomial errors. We see excellent agreement between the four simulations plotted.}  \label{fig:tidal_feature_fraction_halo_mass}
\end{figure*}

\begin{figure*}
    \centering
    \includegraphics[width=\linewidth]{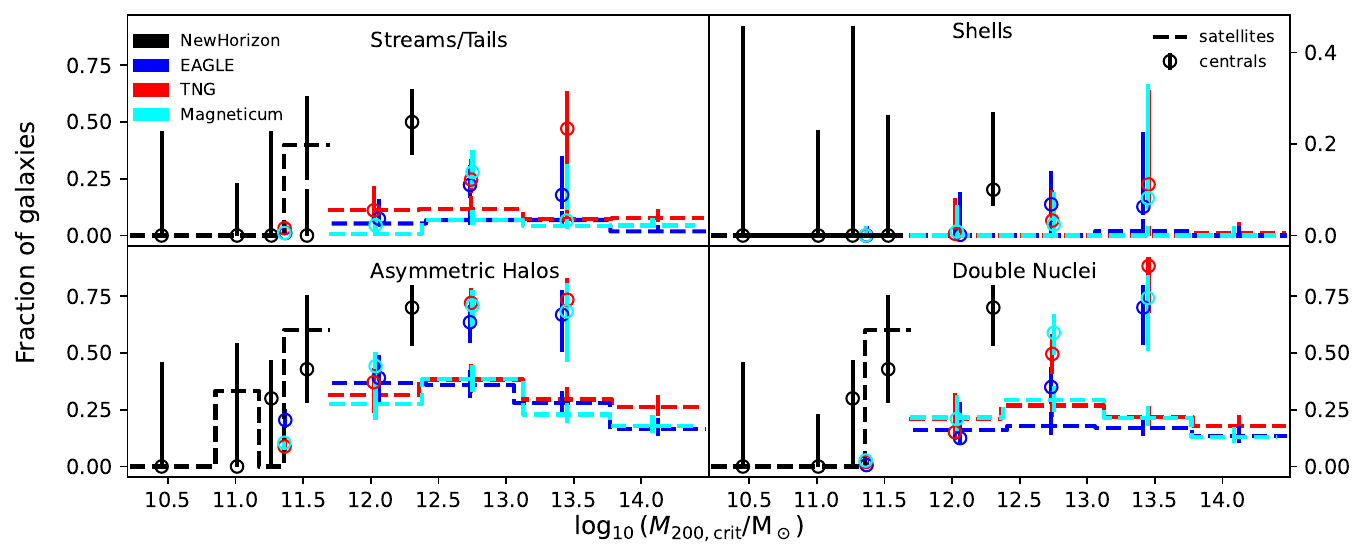}
    \caption{Distribution of tidal feature fractions for satellite and central galaxies as a function of halo mass. From left to right, top to bottom, we have the: stream/tail, shell, asymmetric halo, and double nucleus fractions. The results for \textsc{NewHorizon} are given in black, \textsc{EAGLE} in blue, \textsc{TNG} in red, and \textsc{Magneticum} in cyan. The \textsc{NewHorizon} data is given for the sample of 58 classified galaxies with assigned parent halos. The dashed lines show the feature fractions for satellite galaxies and the circles show the feature fractions for centrals for each halo mass bin containing $\simeq10$ galaxies. For the Monte Carlo sampled distributions the points for centrals and satellites show the means of 100 iterations, each bin contains 90 galaxies. Each bin contains a varying mix of central and satellite galaxies as can be seen in Figure \ref{fig:sample_halo_mass_distribution}. The error bars for \textsc{NewHorizon} shows the $1\sigma$ binomial errors and the error bars for \textsc{EAGLE}, \textsc{TNG} and \textsc{Magneticum} data show the mean $1\sigma$ binomial errors.}
    \label{fig:tidal_feature_fraction_halo_mass_cent_sat}
\end{figure*}

The left panel of Figure \ref{fig:total_tidal_feature_fraction_halo_mass} shows our total tidal feature fractions as a function of halo mass for each of our simulations. We again find excellent agreement between our simulations, with \textsc{EAGLE}, \textsc{TNG} and \textsc{Magneticum} following the same trends and having consistent values for tidal feature fractions in each halo mass bin. For these simulations the fractions of galaxies hosting tidal features increase from halo masses of $10^{11}$ to $\sim10^{12.7}$ M$_{\scriptstyle\odot}$ but decreases with increasing halo mass for halo masses above this. \textsc{NewHorizon} follows a similar trend for the halo masses it covers, with systematically higher fractions.

By comparing the location of the peak in tidal feature fractions with halo mass in the left panel of Figure \ref{fig:total_tidal_feature_fraction_halo_mass} to the proportions of central and satellite galaxies in our sample in Figure \ref{fig:sample_halo_mass_distribution}, we see the peak coincides with the transition from a central-dominated population to a satellite-dominated population. Therefore, we further analyse this trend by comparing the total tidal feature fractions for central and satellite galaxies as a function of the parent halo mass in the right panel of Figure \ref{fig:total_tidal_feature_fraction_halo_mass}. We see again similarly excellent agreement between the simulations, with \textsc{EAGLE}, \textsc{TNG} and \textsc{Magneticum} coinciding in their fractions for both central and satellite galaxies and \textsc{NewHorizon} following similar trends for centrals although at a systematically higher fraction. The fraction of central galaxies exhibiting tidal features increases with increasing halo mass. We note that there is substantial noise in the measurement in the $\sim10^{13.5}$ M$_{\scriptstyle\odot}$ halo mass bin as it contains on average $\lesssim10$ galaxies. The fractions of central galaxies exhibiting tidal features increase to fractions of $0.86^{+0.05}_{-0.2}$, $1.0_{-0.2}$ and $0.73^{+0.1}_{-0.2}$ by $M_{200,\text{ crit}}\gtrsim10^{13}$ M$_{\scriptstyle \odot}$ for \textsc{EAGLE}, \textsc{TNG} and \textsc{Magneticum}. This indicates that at cluster-like halo masses a majority of central galaxies exhibit visually identifiable tidal features. 

The satellite populations for \textsc{EAGLE}, \textsc{TNG} and \textsc{Magneticum} show a declining trend in tidal feature fraction with increasing halo mass. There are only 11 satellites in \textsc{NewHorizon} so it is difficult to comment on any trends with halo mass, but for the halo mass bin centred around $10^{11.3}$ M$_{\scriptstyle\odot}$ we find a significantly higher fraction than the closest halo mass bins for the other three simulations, in line with the higher tidal feature fractions we find for \textsc{NewHorizon} throughout.

Figure \ref{fig:tidal_feature_fraction_halo_mass} shows our specific tidal feature fractions as a function of halo mass. The top left panel shows the stream/tail fraction as a function of halo mass. We find good agreement between the mean distributions for \textsc{EAGLE}, \textsc{TNG} and \textsc{Magneticum}. The fractions of galaxies with features appear to increase from a halo mass of $10^{11}$ M$_{\scriptstyle \odot}$ to $\sim10^{12.7}$ M$_{\scriptstyle \odot}$, from a fraction of $\sim0$ to $\sim0.15$ - $0.2$, before decreasing again down to a fraction of $\sim0$ - $0.05$.  \textsc{NewHorizon} follows a similar shape for the range of halo masses it covers, however, with systematically higher fractions. 

The top right panel of Figure \ref{fig:tidal_feature_fraction_halo_mass} shows the shell fraction as a function of halo mass. While there are very few shells in the sample, there is a small peak in the shell fraction at a halo mass $\sim10^{12.7}$ M$_{\scriptstyle \odot}$. The single shell identified in the \textsc{NewHorizon} simulation does not allow for any meaningful comment on any potential trend. The peak is at a similar halo mass as for the other tidal features. 

The fraction of asymmetric halos with halo mass is presented in the bottom right panel of Figure \ref{fig:tidal_feature_fraction_halo_mass}. The shape resembles that seen for streams/tails and shells, albeit the peak is more clearly resolved due to the larger number of asymmetric halos. The peak is at a similar halo mass ($\sim10^{12.7}$ M$_{\scriptstyle \odot}$) to the peak for streams/tails and shells, and the fraction is larger in size (maximum asymmetric halo fraction $\sim0.5$). \textsc{NewHorizon} appears to again have a systematically higher fraction for each halo mass, whilst following the same trend for the halo masses it covers.

Lastly, the bottom right panel presents the double nucleus fraction as a function of halo mass. We again see a steady increase, peak and decrease in feature fraction, similar to the other tidal feature morphologies. The halo mass peak is consistent with the peak found for the other features. \textsc{NewHorizon} follows a similar trend for the halo masses it covers with a systematically higher double nuclei fraction.

Generally, we see a trend of increasing tidal feature fraction with halo mass to $M_{\scriptstyle200,\text{ crit}}\sim10^{12.7}$ M$_{\scriptstyle \odot}$, after which the fraction appears to decrease. \textsc{NewHorizon} appears to resemble the distributions of the other simulations for the range of halo masses it covers, despite the lack of mass matching and much smaller sample size. 

Figure \ref{fig:tidal_feature_fraction_halo_mass_cent_sat} shows the specific tidal feature fractions for the central and satellite galaxy populations. For each of the feature types, we see that the tidal feature fractions of central galaxies tend to increase with increasing halo mass for each simulation, with the exception of the stream/tail fraction in the halo mass bin centred around $\sim10^{13.5}$ M$_{\scriptstyle\odot}$ for \textsc{EAGLE} and \textsc{Magneticum}. This could be a consequence of the small number statistics of central galaxies in this bin, on average there are only $10.21^{+0.04}_{-0.03}$ and $5.54^{+0.04}_{-0.02}$ centrals in that halo mass bin for \textsc{EAGLE} and \textsc{Magneticum} respectively. We find that shells appear almost exclusively in central galaxies in all the simulations.

For tails/streams in Figure \ref{fig:tidal_feature_fraction_halo_mass_cent_sat}, satellite galaxies appear to follow a flat distribution with halo mass. For asymmetric halos we find that there is some evidence for a small peak in the distribution at $M_{200, \rm crit}\sim10^{12.7}$ M$_{\scriptstyle \odot}$. We see similar evidence for a peak in the distributions of double nuclei fractions in the satellite populations of \textsc{TNG} and \textsc{Magneticum}. However, the distribution of the fraction of satellites exhibiting double nuclei from \textsc{EAGLE} remains flat with respect to halo mass.

\section{Discussion}
\label{sec:Discussion}

We have presented a visual classification of the tidal features around galaxies in four cosmological hydrodynamical simulations: \textsc{NewHorizon}, \textsc{EAGLE}, \textsc{TNG} and \textsc{Magneticum}. Our visual classification finds broad agreement between the four simulations regarding the tidal feature fractions and their behaviour with stellar mass and halo mass, while showing systematically higher tidal feature fractions for \textsc{NewHorizon} in the moderate and highest confidence levels. In \S\ref{subsec:simulations_comparison} we compare our results to previous studies of tidal features in simulations and in \S\ref{subsec:observations} we compare them to observational results to ensure that they are consistent with previous work. In \S\ref{subsec:discussion_visual_classification}, we discuss the impact of our visual classification method, namely having the classifications done by an individual rather than a group. In \S\ref{subsec:differences}, we discuss the possible implications of our results on the occurrence of tidal features in cosmological hydrodynamical simulations and in \S\ref{subsec:environment} we explore the potential of an environmental dependence for tidal feature occurrence in our results. We go on to give testable predictions for LSST in \S\ref{subsec:lsst_predictions}.

\subsection{Comparison with other simulated analyses}
\label{subsec:simulations_comparison}

\begin{figure}
    \centering
    \includegraphics[width=\linewidth]{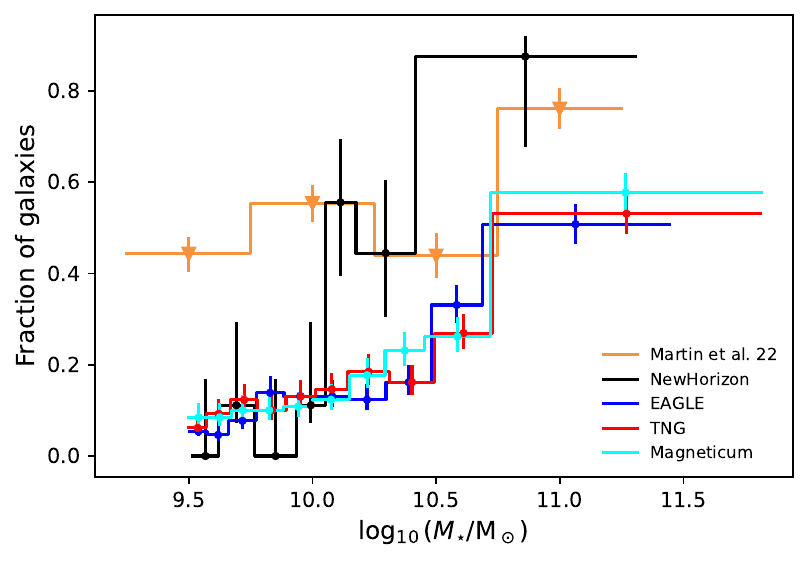}
    \caption{Distribution with stellar mass of the total tidal feature fraction (excluding asymmetric halos) presented with results from \citet{martinPreparingLowSurface2022}. Stellar mass is measured by $M_{\scriptstyle\star,\text{ 30 pkpc}}$ for our sample and \textsc{AdaptaHOP} stellar masses for \citet{martinPreparingLowSurface2022}. \textsc{NewHorizon} is shown in black, \textsc{EAGLE} in blue, \textsc{TNG} in red and \textsc{Magneticum} in cyan. The \citet{martinPreparingLowSurface2022} results are shown in orange. We compute the fractions using classifications with $conf\geq2$. The error bars indicate the $1\sigma$ binomial uncertainty for our results and the $1\sigma$ uncertainty obtained from 100 000 bootstraps for \citet{martinPreparingLowSurface2022}. For each simulation, each mass bin contains a similar number of galaxies, $N_{\mathrm{bin},\textsc{NewHorizon}}\simeq9$ and $N_{\mathrm{bin},\textsc{EAGLE}}=N_{\mathrm{bin},\textsc{TNG}}=N_{\mathrm{bin},\textsc{Magneticum}}=130$. \citet{martinPreparingLowSurface2022} bin their galaxies using stellar mass bins of equal width.}
    \label{fig:all_features_mstar}
\end{figure}

We check whether our results are consistent with previous analyses of tidal features around galaxies in simulations. \citet{martinPreparingLowSurface2022} performed a study of visually-classified tidal features using \textsc{NewHorizon} across a similar range of stellar masses to our study. Their original sample consisted of 37 galaxies from the $z=0.2$ timestep and their progenitors at $z=0.4, 0.6$ and $0.8$, leading to a sample of 148 galaxies. They had 45 astronomers visually classify the images and varied galaxy orientations relative to the observer and the distances at which the object was placed relative to the observer in order to gather $\sim8000$ images, 600 of which were classified by up to 5 different classifiers. They applied an extended classification scheme, including bridges and merger remnants in addition to the classifications we use here. Comparing our results for the fractions of galaxies exhibiting at least one instance of a tidal feature morphology (Figure \ref{fig:tidal_feature_fraction_mass}), we see qualitatively similar trends of the fraction of galaxies exhibiting streams/tails, double nuclei and asymmetric halos all increasing with increasing stellar mass. 

The most easily comparable measure between the two studies is the total tidal feature fraction. We compare to the \citet{martinPreparingLowSurface2022} results for all tidal features, excluding asymmetric halos to align with their approach, in Figure \ref{fig:all_features_mstar}. The results are depicted similarly to Figure \ref{fig:tidal_feature_fraction_mass}, with identical binning and including only classifications with $conf.\geq2$. Our \textsc{NewHorizon} results agree well for stellar masses $\geq10^{10}$ M$_{\scriptstyle \odot}$. For stellar masses below $10^{10}$ M$_{\scriptstyle \odot}$, we report lower tidal feature fractions, with our fractions being lower by $\sim0.4$. For galaxies with $10^{9.5}$ M$_{\scriptstyle \odot}\leq M_{\scriptstyle \star}\leq10^{10}$ M$_{\scriptstyle \odot}$, \citet{martinPreparingLowSurface2022} sample 7 galaxies from the $z=0.2$ timestep, excluding the remaining galaxies due to their halos being contaminated by lower resolution dark matter particles. This contamination is exclusively a concern for zoom-in simulations and therefore only impacts \textsc{NewHorizon} in our sample. We include these galaxies, sampling 32 galaxies in this range to maximise our number statistics for \textsc{NewHorizon}. This difference in sampling is the likely cause of the differences in the measured fraction.

We further check for consistency by comparing our results to those from \citet{valenzuelaStreamComeTrue2023}. They visually classified the presence of streams, tails and shells in the same \textsc{Magneticum} simulation box we use for this study. \citet{valenzuelaStreamComeTrue2023} performed their visual classification on the 3D stellar component of \textsc{Magneticum} galaxies with $M_{\scriptstyle\star}\geq2.4\times10^{10}$ M$_{\scriptstyle \odot}$ and virial mass $M_{\rm vir}\geq7.1\times10^{11}$ M$_{\scriptstyle \odot}$ ($M_{\rm vir}\approx M_{200, \rm \:crit}$). The  most direct comparison to our results is using the total tidal feature fractions. In Figure \ref{fig:tidal_feature_fractions_comparison}, we show the fraction of galaxies exhibiting tidal features in each of our samples, for $conf.\geq1$ and $conf.=3$ classifications. The confidence levels increase with the opacity of the point. The left-hand panel compares the simulated tidal feature fractions. The stellar mass error bars for each of the fractions give the range of stellar masses studied (using the $M_{\star,\text{ 30 pkpc}}$ masses for our results), while the point itself gives the fraction and the mean stellar mass where the measurement is available, and the midpoint of the stellar mass range where the mean is unavailable. \citet{valenzuelaStreamComeTrue2023} measured a tidal feature fraction of $0.23\pm0.02$, which agrees well with our tidal feature fraction for \textsc{Magneticum} and other simulations at $conf.=3$. Limiting ourselves to just tails, streams and shells within the stellar mass range covered by \citet{valenzuelaStreamComeTrue2023}, we find a feature fraction of $0.17\pm0.02$ at $conf.\geq1$ and $0.09^{+0.02}_{-0.01}$ for $conf.=3$. Therefore, our fractions sit  a factor of $\sim1.4-2.5$ times below the fractions for the same \textsc{Magneticum} simulation box. \citet{valenzuelaStreamComeTrue2023} also find an increasing fraction of galaxies exhibiting tidal features with stellar mass, qualitatively agreeing with our predictions. The differences in our fractions could be due to us visually detecting and classifying tidal features from mock images while \citet{valenzuelaStreamComeTrue2023}  classified based on unprojected 3D data. Mock images are subject to projection effects and the surface brightness limit we have applied which \citet{popFormationIncidenceShell2018,kado-fongTidalFeatures052018} and \citet{martinPreparingLowSurface2022} suggest have a significant impact on the detectability of tidal features. Therefore, lower fractions from these projected mock images are expected.

\citet{popFormationIncidenceShell2018} studied shell formation in an analogous run used from the previous generation \textsc{Illustris} simulation which has a similar resolution to that used in this study \citep{genelIntroducingIllustrisProject2014, vogelsbergerIntroducingIllustrisProject2014}. They performed their analysis on 220 of the most massive galaxies in the \textsc{Illustris} simulation ($M_{\scriptstyle \star}\gtrsim10^{11}$M$_{\scriptstyle \odot}$) and visually detected the presence of shells using stellar surface density maps across 3 orthogonal projections. Their shell fraction of $0.18^{+0.03}_{-0.02}$ is presented in the left panel of Figure \ref{fig:tidal_feature_fractions_comparison}. Furthermore, \citet{popFormationIncidenceShell2018} predict that in an observational survey (i.e. with only one projection available) they would measure a shell fraction of $0.14\pm0.03$. For a direct comparison, we take our most optimistic shell fraction ($conf.\geq1$) for our stellar mass-matched \textsc{TNG} sample. This gives a substantially lower fraction of shells than \citet{popFormationIncidenceShell2018}, $0.005^{+0.003}_{-0.001}$ (Table \ref{tab:tidal_feature_fraction}). From Figure \ref{fig:tidal_feature_fractions_comparison}, we can see that our studies cover different mass ranges. \citet{popFormationIncidenceShell2018} probes galaxies that are much higher in stellar mass than ours. As shell fractions increase with stellar mass, we find it likely that the different stellar mass ranges of the samples contribute to the different tidal feature fractions. To test the impact of the different stellar mass ranges, we extrapolate our shell fraction to the midpoint of \citet{popFormationIncidenceShell2018}'s stellar mass range, $10^{12.3}$ M$_{\scriptstyle \odot}$. While our $conf.\geq2$ results plotted in Figure \ref{fig:tidal_feature_fraction_mass} only have measured shell fractions in the highest stellar mass bin, if we include $conf.\geq1$ classifications we obtain measurements in the last two stellar mass bins of $0.008^{+0.02}_{-0.002}$ and $0.05^{+0.03}_{-0.01}$. An approximate linear gradient of $0.1\pm0.1$ can be obtained using the mean galaxy stellar masses in these two bins. Using this as an estimate for the rate of shell fraction increase with stellar mass for our sample, we can estimate what our total shell fraction would be if our stellar masses are extrapolated from the mean mass in our highest stellar mass bin to the midpoint of \citet{popFormationIncidenceShell2018} stellar mass range. This gives a shell fraction $0.18^{+0.17}_{-0.15}$, which is consistent with the fraction measured by \citet{popFormationIncidenceShell2018}, suggesting that the differences are mainly due to the stellar mass ranges of the respective samples. 

Figure \ref{fig:tidal_feature_fraction_halo_mass} shows the relationship between halo mass and tidal feature occurrence. It suggests that we may be able to test the rates of close encounters as a function of the environment predicted by N-body simulations \citep[e.g.][]{gnedinTidalEffectsClusters2003} using observations from LSST. The simulations predicted that the rate of close encounters per galaxy as a function of the relative velocities of the galaxies during the encounter has a peak at an encounter velocity $\sim500$ km s$^{-1}$. This corresponds to 2 galaxies moving at speeds $\sim250$ km s$^{-1}$ relative to the centre of mass of the cluster. The halo mass-velocity dispersion relationship shows that $\sigma\simeq250$ km s$^{-1}$ corresponds to a peak at $M_{\rm vir}\sim10^{13}$ M$_{\scriptstyle \odot}$ \citep{elahiUsingVelocityDispersion2018}. We expect there to be a corresponding peak in the relationship we measure between tidal feature occurrence and halo mass. As tidal features are tracers of galaxy encounters, the fact that we see a peak at $M_{200, \rm crit}\sim10^{12.7}$ M$_{\scriptstyle \odot}$ in Figure \ref{fig:tidal_feature_fraction_halo_mass} is an encouraging prediction from our simulations. 

\citet{jianEnvironmentalDependenceGalaxy2012} studied close pairs in the \textsc{Millenium} N-body simulation and a suite of semi-analytic models. They found a peak in the relationship between the fractions of close pairs undergoing mergers and halo mass between $10^{12}$ - $10^{13}$ M$_{\scriptstyle \odot}$. This is in good agreement with our double nuclei panel in Figure \ref{fig:tidal_feature_fraction_halo_mass}.

We find good agreement between the tidal feature fractions in our simulations, suggesting that tidal features are a genuine probe of mass assembly and may not be sensitive to processes being modeled by subgrid physics. \citet{canasStellarHaloesIntracluster2020} and \citet{proctorIdentifyingDiscBulge2023} studied intra-halo light (IHL) in \textsc{Horizon-AGN} and \textsc{EAGLE} respectively and \citet{broughPreparingLowSurface2024} studied the IHL in \textsc{Horizon-AGN}, \textsc{Hydrangea}, \textsc{Magneticum} and \textsc{IllustrisTNG}. These works all found good agreement in IHL fractions despite the different subgrid physics models between simulations. As IHL is a similarly low surface brightness feature formed through mergers, the agreement between IHL fractions across these simulations suggests that gravitational physics is playing a dominant role in this probe of assembly history.

In general, we find an encouraging agreement with existing simulation results. The differences in the tidal feature fractions in our results and existing studies are likely explained by a combination of the differences in stellar mass ranges and the potential reduction of tidal feature detectability due to the surface brightness limits of the mock images.

\subsection{Comparison with observational results}
\label{subsec:observations}

\begin{figure*}
    \centering
    \includegraphics[width=\linewidth]{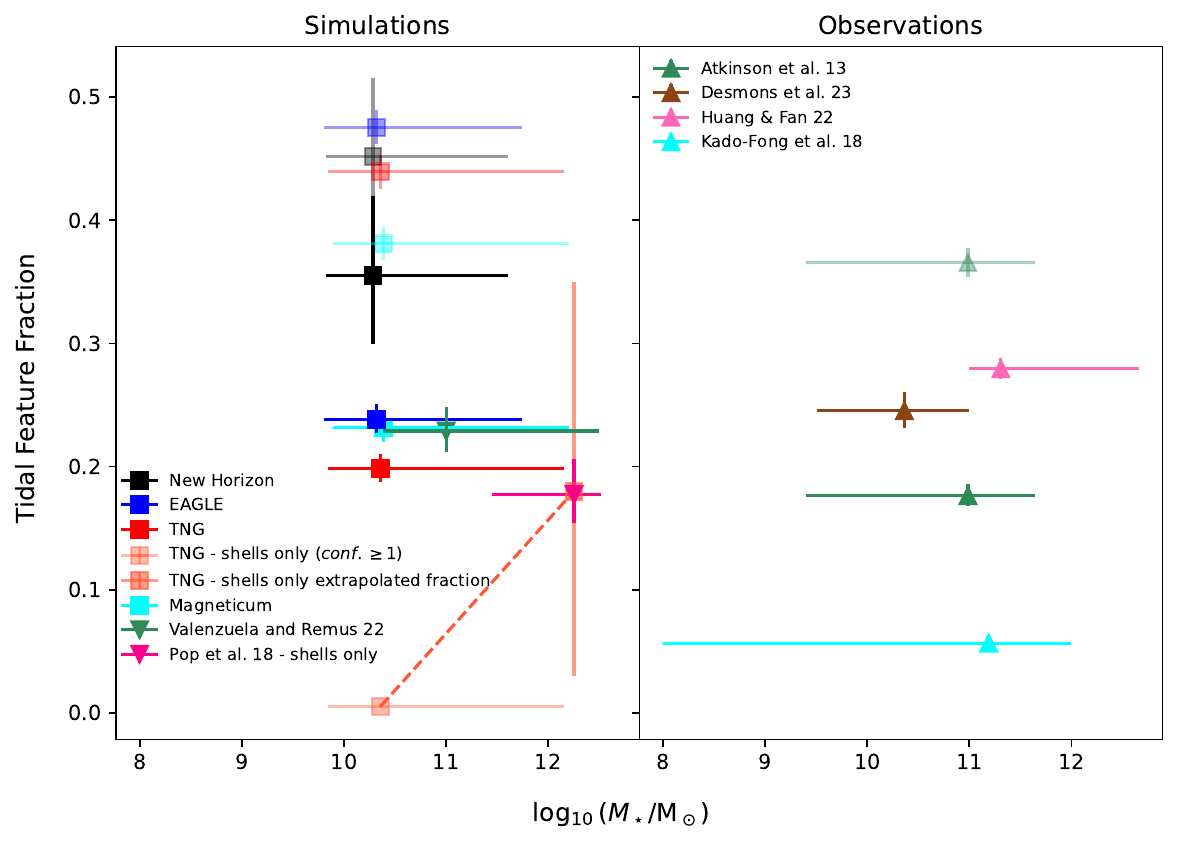}
    \caption{Total fraction of galaxies exhibiting tidal features from various observations and simulations studies compared to the results presented here. On the left, we display our own simulation results and simulation results from the literature. On the right, we give the results of observational studies. The error bars along the stellar mass axis show the range of stellar masses for each measurement, with the marker situated on the mean where the sample mean stellar mass is provided (where unavailable the midpoint of the stellar mass range is used instead). Our $conf.\geq1$ results are presented with a transparent marker and our $conf.=3$ results are presented with the solid marker. We use a similar opacity to indicate the different confidence level derived fractions from observations. The dashed line links our TNG $conf.\geq1$ shell fraction with an estimated fraction when linearly extrapolated to higher stellar masses. We find our tidal feature fractions to be in the expected range for observational results and simulation results.}
    \label{fig:tidal_feature_fractions_comparison}
\end{figure*}

\begin{figure*}
    \centering
    \includegraphics[width=\linewidth]{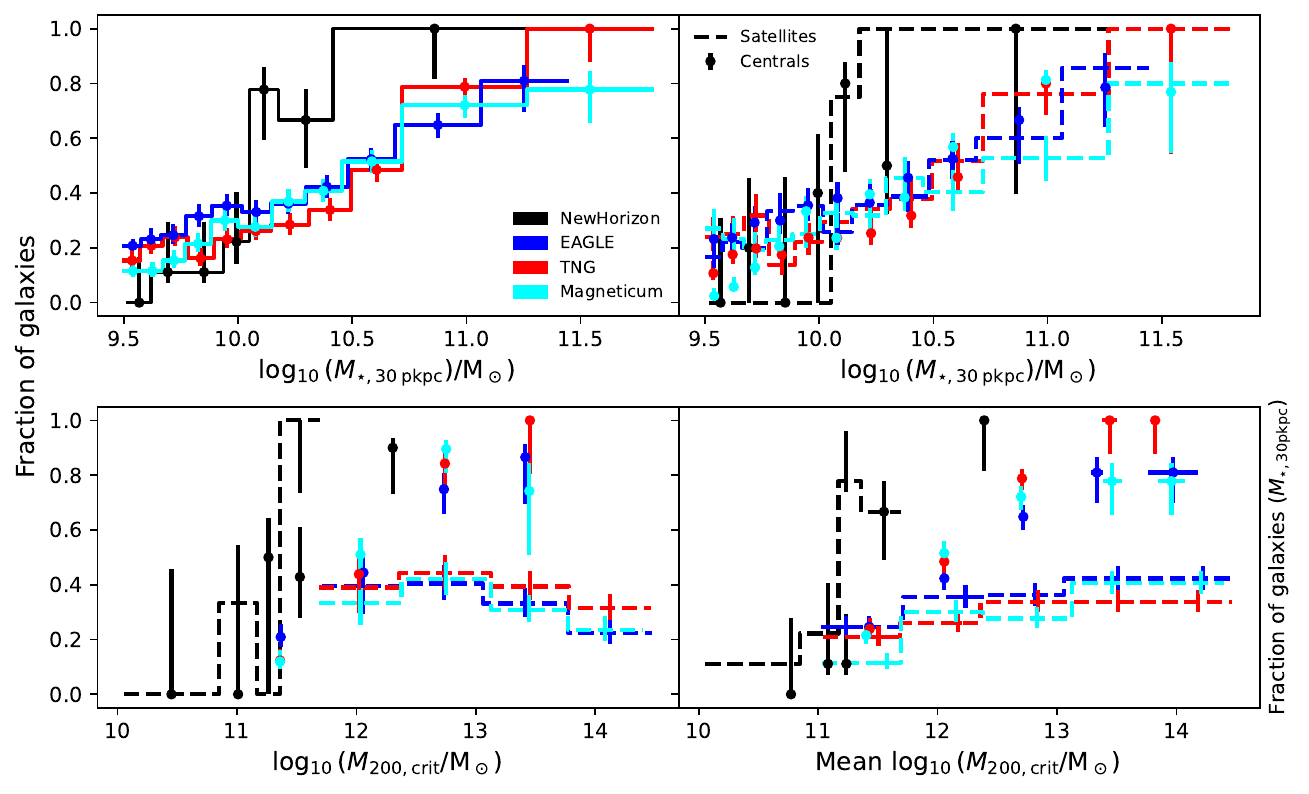}
    \caption{Investigating whether trends of tidal feature fraction with halo mass and as a function of central and satellite galaxy populations can be explained by the correlation between tidal feature fraction and galaxy stellar mass. The top left panel shows the tidal feature fraction-stellar mass relation shown in Figure \ref{fig:all_features_mstar} with the inclusion of asymmetric halos and with the final bin for \textsc{EAGLE}, \textsc{TNG} and \textsc{Magneticum} divided into two bins, linearly spaced in log-stellar mass. The top right panel uses identical binning to the top left, and plots the fractions of centrals and satellite galaxies exhibiting tidal features as a function of stellar mass. This illustrates that there are no differences in the relations followed by central or satellite galaxies across all simulations. The bottom left panel is identical to the right panel of Figure \ref{fig:total_tidal_feature_fraction_halo_mass} and shows the tidal feature fraction as a function of halo mass for central and satellite galaxies. The bottom right panel plots the tidal feature fraction estimated from the tidal feature fraction-stellar mass relation shown in the top left panel, for the populations of centrals and satellites as a function of mean halo mass. For central galaxies we find reasonable agreement between the bottom two panels, indicating that the trends seen in the tidal feature fraction-halo mass relation are a result of the relationship of the tidal feature fractions with stellar mass. However, satellites in the bottom right panel have fractions that systematically increase with increasing halo mass compared to the bottom left panel, where they increase and then fall as a function of increasing halo mass. This could indicate that there are factors other than stellar mass that drive tidal feature fractions for satellite galaxies. The uncertainties in the tidal feature fraction are given by the $1\sigma$ binomial errors for the relations with stellar mass and the mean $1\sigma$ binomial errors for the relations with halo mass. The uncertainties on the mean halo mass are given by standard deviation of the mean halo mass over the 100 Monte Carlo iterations.}
    \label{fig:stellar_mass_trend_explains_halo_mass}
\end{figure*}

Observed tidal feature fractions measured by prior studies span a large range of values. Most of these differences are likely driven by the different surface brightness limits of the surveys, as tidal features become more visible with deeper imaging and at higher galaxy stellar masses \citep[e.g.][]{kado-fongTidalFeatures052018, martinPreparingLowSurface2022}. To ease comparison with the values measured here we have plotted the total tidal feature fractions measured in observational studies in the right-hand panel of Figure \ref{fig:tidal_feature_fractions_comparison}.

\citet{atkinsonFAINTTIDALFEATURES2013} used observations from the wide component of the Canada-France-Hawaii Telescope Legacy Survey. They visually classified \textit{g'-r'-i'} stacked images of 1781 galaxies with a redshift range of $0.02<z<0.4$. The limiting surface brightness of their data was $\mu_{g'_{\text{AB}}}=27.7\pm0.5$ mag arcsec$^{-2}$. The classification scheme used by \citet{atkinsonFAINTTIDALFEATURES2013} identifies; streams, arms, linear features, miscellaneous structure, diffuse fans and shells and has significant overlap with our own classification types. Our streams/tails category is roughly analogous to their streams $+$ arms $+$ linear features, and our asymmetric halo category corresponds to their diffuse fans + miscellaneous structure. For their lowest confidence level \citet{atkinsonFAINTTIDALFEATURES2013} measure a total tidal feature fraction of $0.37\pm0.01$ and at their highest confidence level, they measure $0.18\pm0.01$. Their lowest confidence level sits below our $conf.\geq1$ fractions for \textsc{NewHorizon}, \textsc{EAGLE} and \textsc{TNG}: $0.45\pm0.06$, $0.48\pm0.01$ and $0.44\pm0.01$. However, their fraction is comparable to \textsc{Magneticum}: $0.38\pm0.01$, which presents a tidal feature fraction slightly lower than the other simulations, likely due to its slightly lower resolution discussed in more detail in \S\ref{subsec:differences}. Their highest confidence level tidal feature fraction sits significantly below our $conf.=3$ fractions for \textsc{NewHorizon}, \textsc{EAGLE}, \textsc{TNG} and \textsc{Magneticum}: $0.35\pm0.06$, $0.24\pm0.01$, $0.20\pm0.01$ and $0.23\pm0.01$. This is expected given \citet{atkinsonFAINTTIDALFEATURES2013} examined data at a limiting surface brightness $\sim2.6$ mag arcsec$^{-2}$ brighter than our mock images.

There have been many studies of tidal features made using HSC-SSP observations, such as the work of \citet{kado-fongTidalFeatures052018}, \citet{huangMassiveEarlyTypeGalaxies2022} and \citet{desmonsGalaxyMassAssembly2023}. \citet{kado-fongTidalFeatures052018} studied 20,000 galaxies across a redshift range of $0.05<z<0.45$ using the first data release of the HSC-Wide catalogue \citep{boschHyperSuprimeCamSoftware2018}. They used an automated method, creating high spatial frequency images to detect tidal features and then visually classifying the morphologies of these features. They found a much lower total tidal feature fraction of $0.056\pm0.002$. They cover a much larger range of stellar masses than our sample and detect tidal features to surface brightness depths of $\mu_{i}\sim26.4$ mag arcsec$^{-2}$, 3.3 mag arcsec$^{-2}$ shallower than our mock images. With their stellar mass range probing down to $10^{8}$ M$_{\scriptstyle \odot}$ and redshift range reaching $z\sim0.45$, many tidal features will fall below their detection limit resulting in their lower fraction.

\citet{huangMassiveEarlyTypeGalaxies2022} used the coadded images from the Wide layer of HSC-SSP third data release, which reaches a surface brightness limit of $\mu_{i}\sim28.5$ mag arcsec$^{-2}$ ($\sim1.2$ mag arcsec$^{-2}$ brighter than our mocks). Their final sample of galaxies consists of 2649 early-type galaxies without companions (i.e. the sample has no double nuclei) with stellar masses $\geq10^{11}$ M$_{\scriptstyle \odot}$. They apply an automated procedure to model and remove the host galaxy and leave behind only the tidal features. Their total tidal feature fraction is $0.28\pm0.01$. Their fraction is slightly higher than our $conf.=3$ fractions for all of our simulations other than \textsc{NewHorizon}. Their fraction falls $\sim0.2$ below our $conf.=1$ fractions, this indicates that when we include all our galaxies that exhibit any trace of a tidal disturbance, even where it is difficult to determine its morphology, we visually detect far more than \citet{huangMassiveEarlyTypeGalaxies2022}. This is likely due to the deeper surface brightness limits applied to our images. 

\citet{desmonsGalaxyMassAssembly2023} used the HSC-SSP second data release UltraDeep region images, with a limiting surface brightness of $\mu_{g}\sim30.76$, $\mu_{r}\sim29.82$ and $\mu_{i}\sim29.41$ mag arcsec$^{-2}$ ($3\sigma$; $10''\times10''$). They limited their sample to galaxies in the redshift range $0.04<z<0.2$ and applied stellar mass limits of $10^{9.5}$ M$_{\scriptstyle \odot}$$\leq M_{\scriptstyle \star}\leq10^{11}$ M$_{\scriptstyle \odot}$, constructing a volume-limited sample of 852 galaxies. Their classification scheme is identical to ours, making this a particularly relevant comparison. They measure a total tidal feature fraction of $0.23\pm0.02$. This is in good agreement with our $conf.=3$ results for \textsc{EAGLE}, \textsc{TNG} and \textsc{Magneticum} and $\sim0.1$ lower in fraction than \textsc{NewHorizon}, while we measure significantly higher fractions when including all possible tidal features ($conf.\geq1$). Given their lower stellar mass range, it could be that the differences in fractions are driven by the contribution of simulated galaxies with $M_{\scriptstyle\star, \text{ 30 pkpc}}\geq10^{11}$ M$_{\scriptstyle\odot}$ that tend to have higher tidal feature fractions than lower stellar mass galaxies. Furthermore, all the galaxies in the \citet{desmonsGalaxyMassAssembly2023} sample, are further from the observer than in our sample. Galaxies at $z=0.2$ are further away than the galaxies in our mock images, more distant tidal features may be too dim to detect visually. \citet{desmonsGalaxyMassAssembly2023} found qualitatively similar trends with increasing tidal feature fractions as a function of stellar mass. They do detect a substantially higher fraction of galaxies exhibiting shells, $0.02\pm0.01$, higher than our mean $conf.\geq1$ shell fraction of $0.006^{+0.001}_{-0.0007}$ (excluding \textsc{NewHorizon} as we only detect one shell in this simulation). This result is surprising as the \citet{desmonsGalaxyMassAssembly2023} sample is measured over lower stellar masses than ours. Given that shell fraction increases with increasing galaxy stellar mass we would expect our sample to return a higher shell fraction. Constraining our sample to $M_{\scriptstyle\star, \text{ 30 pkpc}}\leq10^{11}$ M$_{\odot}$, we find \textsc{EAGLE}, \textsc{TNG} and \textsc{Magneticum} to have mean $conf.\geq1$ shell host masses of $\log_{10}(M_{\scriptstyle\star, \text{ 30 pkpc}}/$M$_{\scriptstyle\odot})=$ $10.8\pm0.2$, $10.7\pm0.1$ and $10.8\pm0.1$. These are consistent within 2$\sigma$ of the mean shell host stellar mass of $\log_{10}(M_{\scriptstyle\star}/$M$_{\scriptstyle\odot})=10.56\pm0.04$, found by \citet{desmonsGalaxyMassAssembly2023}.

The higher fractions of galaxies hosting shells in observations could be a result of factors relating to the properties of the host galaxies, the characteristics of the shells themselves and the methods we use to produce the mock images. The detectability of shells was examined in \citet[][]{martinPreparingLowSurface2022} and \citet{bazkiaeiDetectionLimitsGalaxy2023}, who found that it depends significantly on the degree of contrast against the background galaxy. The variation in detectability is driven more by the physical parameters of a galaxy (stellar mass, Sersic index and galaxy size) than by the characteristics of the shells themselves (opening angle and shell width). \citet{bazkiaeiDetectionLimitsGalaxy2023} find that detection of shells is dependent on the stellar mass resolution of the simulations, which limits the amount of contrast a shell can achieve against the host galaxy in the inner radii close to the galaxy. \citet{martinPreparingLowSurface2022} found that within $\sim4$ $R_{\scriptstyle\rm eff.}$, shells will not be detectable in \textsc{NewHorizon} mock images but could be detectable in LSST. We expect this region to extend to larger radii for the other simulations as they have lower stellar mass resolutions \citep{bazkiaeiDetectionLimitsGalaxy2023}. 

It is possible that systematics introduced by the mock image construction (e.g. smoothing) could reduce the visibility of some shells, making shells more difficult to detect or potentially be misidentified as a different class (e.g. asymmetric halo). However, the agreement of our \textsc{TNG} shell fractions with the shell fractions measured directly from \textsc{Illustris} simulation data by \citet{popFormationIncidenceShell2018}, suggests that the mock image creation method does not significantly impact the visibility of shells. Direct comparison with LSST data could help clarify this tension between simulation and observed shell fractions.

\subsection{Discussion of visual classification methodology}
\label{subsec:discussion_visual_classification}

The majority of the visual classification was performed by the lead author. There was consultation on a subset of 31 \textsc{NewHorizon} galaxies to ensure good agreement on the visually classified morphological types of features. Other studies have already used visual classifications by groups of scientists and reported on the stability and convergence of such classification systems. \citet{bilekCensusClassificationLowsurfacebrightness2020} used a group of scientists of different experience levels to identify streams, shells and tails on the MATLAS Survey data. They found that there is a significant scatter between classifiers and that this scatter tends to decrease with the experience of the classifiers. \citet{martinPreparingLowSurface2022} also found that with faint tidal features, there was decreased concurrence between classifiers on the nature of the morphologies, increasing the scatter. The above suggests that using multiple classifiers helps to characterise the scatter in the morphology and detectability of the tidal feature and could increase the accuracy and reliability of visual detection and classification. However, we argue that having a single classifier makes for a similarly biased comparison across all of the simulations, as the biases of the single classifier are carried across the simulations as systematic errors, allowing for a fair consensus on whether the simulations agree with each other.

\subsection{Comparison between the different simulations}
\label{subsec:differences}
The broad agreement between the four different simulations analysed here, and the behaviour of these fractions with respect to stellar mass and halo mass suggests that the different subgrid physics and the different hydrodynamics schemes applied are not playing a dominant role in the observability of tidal features. The impact of classification confidence on the feature fractions is presented in Table \ref{tab:confidence_levels} and Figure \ref{fig:tidal_feature_frequency}. Importantly we see that the decrease in the fractions of galaxies exhibiting tidal features with an increasing confidence level is less significant in the \textsc{NewHorizon} simulation than in the other simulations. This implies that the classified tidal features are of a higher confidence and therefore easier to detect in \textsc{NewHorizon} than the other simulations. A contributing factor to the higher confidence tidal features in the \textsc{NewHorizon} simulation could be its $\sim2$ orders of magnitude higher stellar mass resolution compared to the other simulations. For a given number of particles, \textsc{NewHorizon} would be able to resolve tidal features from disrupted galaxies with $\gtrsim100$ times lower stellar mass than the other simulations, which could result in the higher fractions (Appendix \ref{app:detection_limits}). The slightly lower mass resolution for \textsc{Magneticum} when compared to the other simulations could also be contributing to its lower fraction when including $conf\geq1$ tidal features that are likely to include fainter features where the mass resolution could significantly impact its visibility when near a bright host (Appendix \ref{app:detection_limits}). 

The galaxy SMHM relation (Figure \ref{fig:halo_mass_stellar_mass}), points to a different potential source of \textsc{NewHorizon}'s higher tidal feature fractions as a function of stellar and halo mass. In this relation, we see that for a given parent halo mass \textsc{NewHorizon} has on average higher stellar mass galaxies populating the halo compared to the other simulations. This means that when galaxies undergo mergers, \textsc{NewHorizon} will tend to bring more stellar mass into the merger and therefore, is likely to produce more visible tidal features for a galaxy of a particular stellar mass than the other simulations.

These results suggest that the simulations are in reasonable agreement with one another regarding the occurrence of visually-detectable tidal features. The higher resolution and higher galaxy stellar masses for a given halo mass of \textsc{NewHorizon} could be driving the higher fractions we measure.

\subsection{Tidal feature fractions and environment}
\label{subsec:environment}

The relationships between tidal feature fractions and halo mass, plotted in the left hand panel of Figure \ref{fig:total_tidal_feature_fraction_halo_mass} illustrate a peak in the occurrence of tidal features at $M_{\scriptstyle 200,\text{ crit}}\sim 10^{12.7}$M$_{\scriptstyle\odot}$. This peak coincides with the transition from a central-dominated population of galaxies to a satellite-dominated population. These observations together suggest that the the occurrence of tidal features could depend on environment both measured as a function of halo mass and as location in the halo. In Figure \ref{fig:stellar_mass_trend_explains_halo_mass} we investigate whether the trends we measure with halo mass are indicative of an environmental influence on the occurrence of tidal features or whether the observed trends can be reproduced by accounting for the trends of tidal features with galaxy stellar mass.

We start in the top panel by investigating whether there is any separation in the tidal feature-stellar mass relation when dividing it into central and satellite galaxies. The top left panel is a version of Figure \ref{fig:all_features_mstar}, including asymmetric halos and with the highest stellar mass bin in the \textsc{EAGLE}, \textsc{TNG} and \textsc{Magneticum} simulations, split into two bins linearly spaced in log-stellar mass to allow for the trends in the higher stellar mass ranges ($M_{\scriptstyle\star,\text{ 30 pkpc}}\gtrsim5\times10^{10}$ M$_{\scriptstyle\odot}$) to be explored. The top right panel shows tidal feature fractions with galaxy stellar mass for central and satellite galaxies in each of the bins in the top-left panel. We see for all the simulations that the tidal feature fraction with stellar mass does not depend significantly on whether the galaxy is a satellite or a central.

In the lower panels we explore whether the relationship observed with halo mass in Figure \ref{fig:total_tidal_feature_fraction_halo_mass} could be a result of the observed relationship with stellar mass. We repeat Figure \ref{fig:total_tidal_feature_fraction_halo_mass} in the bottom left panel for direct comparison. The bottom right panel shows the tidal feature fraction estimated using the mean stellar masses of the central and satellite galaxies in each of the bins in the bottom left panel. The trends seen for central galaxies in the bottom left panel are reproduced well in the bottom right panel, suggesting that their tidal feature fractions are mostly driven by their stellar mass. The satellite tidal feature fractions inferred from their stellar mass systematically increase with increasing halo mass reaching $0.3$ - $0.4$ for $M_{\scriptstyle 200,\text{ crit}}\gtrsim10^{13}$ M$_{\odot}$. However, the true satellite tidal feature fractions increase with increasing halo mass to $M_{\scriptstyle 200,\text{ crit}}\sim10^{12.7}$ M$_{\scriptstyle\odot}$ and decrease with increasing halo mass beyond this mass. While this is within $2\sigma$ of the uncertainties on these fractions, the systematic nature of the difference suggests there might be factors other than stellar mass driving the relationship between tidal feature fractions and halo mass for satellite galaxies. This might indicate an enhancement of mergers in satellites residing in halos of $M_{\scriptstyle 200,\text{ crit}}\sim10^{12.7}$ M$_{\scriptstyle\odot}$ and a reduction in mergers in satellites residing in halos of $M_{\scriptstyle 200,\text{ crit}}\gtrsim10^{13}$ \citep[e.g.][]{jianEnvironmentalDependenceGalaxy2012}. The work of \citet{omoriGalaxyMergersSubaru2023} provides another example of galaxy mergers rates exhibiting a dependence on environment. \citet{omoriGalaxyMergersSubaru2023} used a deep learning model to detect mergers in HSC-SSP, finding that on scales of 0.5 to 8 Mpc h$^{-1}$, mergers tended to occur in lower density environments, showing that galaxy environment is a significant factor in the occurrence of mergers. The role of environment in the occurrence of galaxy mergers is further highlighted by \citet{sureshkumarGalaxyMergersPrefer2024}. \citet{sureshkumarGalaxyMergersPrefer2024} studied the spatial clustering of merger and non-merger galaxies in the GAMA catalogue and found that galaxy mergers tend to occur in under-dense environments on spatial scales greater than 50 kpc h$^{-1}$.

\subsection{Predictions for LSST}
\label{subsec:lsst_predictions}

Given the broadly good agreement between the simulations regarding tidal feature fractions and their trends with stellar mass and halo mass, we can be confident in the simulations' predictive power for comparison with LSST. We note that \textsc{NewHorizon}'s higher tidal feature fractions in some relations could potentially indicate that the higher mass resolution of this simulation allows it to resolve tidal features from lower stellar mass progenitors. We use a simple model to probe the limiting progenitor mass for tidal features that each simulation can detect in the mock images in Appendix \ref{app:detection_limits}. In \textsc{NewHorizon} we can detect tidal features down to stellar masses of $\sim10^{6}$ M$_{\scriptstyle\odot}$, so long as they are not disrupted over an area $\gtrsim36$ pkpc$^{2}$ and in the other simulations this limit is $\sim10^{8}$ M$_{\odot}$ with the tidal feature being disrupted over an area $\sim2500$ pkpc$^{2}$ so long as the light from the host galaxy is at a similar surface brightness as the feature or fainter within the same region as the tidal feature. This could suggest that the predictions here offer a lower bound on the potential detections in LSST. However, the higher stellar mass for a given halo mass in \textsc{NewHorizon}'s SMHM relation as well as its smaller volume could also contribute to the higher fractions. The fact that the three simulations of a similar resolution are in agreement regarding the tidal feature fractions and their trends with stellar and halo mass, indicates that the simulations are well converged and offer robust predictions down to the lowest stellar mass tidally disrupted galaxy that can be both resolved in the simulation and visibly detected in \textsc{LSST} \citep[e.g.][]{martinPreparingLowSurface2022}.

Given that the simulations offer robust predictions, complete to their lowest stellar mass tidally disrupted galaxy that can be resolved in our mock images, they offer at a minimum a lower bound on tidal feature fractions should LSST be able to probe tidally tidal features below simulation resolution limits which will depend on observational processing \citep{watkinsStrategiesOptimalSky2024}. We inspect Figure \ref{fig:tidal_feature_fraction_mass} and Figure \ref{fig:total_tidal_feature_fraction_halo_mass} to examine what the simulations may predict with respect to visually-detected tidal features in LSST. The increasing tidal feature fraction with stellar mass is commonly observed in both simulations and observations. This trend supports the observational evidence that the visual detectability of tidal features increases with the stellar mass of the host. A comparison of our trends of tidal feature fraction with stellar mass against future LSST observations will enable us to further test whether these predictions are consistent with the real Universe. We can also probe whether the tidal feature fraction relation with halo mass for satellite galaxies has any dependence on environment.

LSST will survey billions of galaxies. To take full advantage of this unprecedented data set we will need automated detection and classification of tidal features. There is ongoing work on automated detections of tidal features \citep[e.g.][]{kado-fongTidalFeatures052018,pearsonIdentifyingGalaxyMergers2019,desmonsDetectingGalaxyTidal2023,omoriGalaxyMergersSubaru2023a}. \citet{desmonsDetectingGalaxyTidal2023} have developed a self-supervised machine-learning method to detect tidal features in HSC-SSP galaxies. Further work to convert these detections into classifications of specific tidal feature morphologies would enable the measurement of trends of total and specific tidal feature fractions with stellar and halo mass over large samples for which visual classification would be prohibitively time-consuming.
 
\section{Conclusions}
\label{sec:conclusions}

We have presented the results of our visual classifications of tidal features around galaxies with $10^{9.5}$ M$_{\scriptstyle\odot}\leq M_{\scriptstyle\star}\lesssim10^{12}$ M$_{\scriptstyle\odot}$ in mock images made to predicted LSST 10-year surface brightness depths of $\mu_{g}\sim30.3$ mag arcsec$^{-2}$, $\mu_{r}\sim30.3$ mag arcsec$^{-2}$ and $\mu_i\sim29.7$ mag arcsec$^{-2}$ from four cosmological hydrodynamical simulations. We compared the tidal feature fractions and their behaviour as a function of stellar and halo mass. We draw the following conclusions from our results:
\begin{itemize}
    \item Tidal feature fractions decrease with increasing classification confidence levels. This highlights the need to specify a confidence level of visual classification appropriate for the analysis. We present our results using $conf.\geq2$ (see Table \ref{tab:confidence_levels}).
    \item The total tidal feature fractions are consistent between the simulations: $f_{\textsc{NewHorizon}}=0.40\pm0.06$, $f_{\textsc{EAGLE}}=0.37\pm0.01$, $f_{\textsc{TNG}}=0.32\pm0.01$ and $f_{\textsc{Magneticum}}=0.32\pm0.01$ (Table \ref{tab:tidal_feature_fraction}). This shows that the occurrence of visually identified tidal features in the cosmological hydrodynamical simulations may not be sensitive to the different subgrid physics applied by each.
    \item The higher tidal feature fractions for \textsc{NewHorizon} as a function of stellar and halo mass may be driven by differences between its galaxy SMHM relation and that of the other simulations (Figure \ref{fig:halo_mass_stellar_mass}) as well as its higher stellar mass resolution (Appendix \ref{app:detection_limits}).
    \item The impact of simulation stellar mass resolution is similar for all feature types, generally enhancing the likelihood of detecting a feature around a given galaxy.
    \item The specific tidal feature fractions are consistent across the four simulations (values for each simulation given in Table \ref{tab:tidal_feature_fraction}), with average specific feature fractions across the four simulations of: $f_{\text{stream/tail}}=0.07^{+0.05}_{-0.03}$, $f_{\text{shell}}=0.0041^{+0.002}_{-0.0007}$, $f_{\text{asym.}}=0.29^{+0.07}_{-0.005}$ and $f_{\text{DN}}=0.17^{+0.07}_{-0.05}$.
    \item Total tidal feature fractions across the simulations increase with increasing stellar mass from $0.0^{+0.2}$, $0.208^{0.02}_{-0.009}$, $0.15^{+0.04}_{-0.03}$ and $0.12^{+0.03}_{-0.02}$ to $\sim1_{-0.2}$, $0.67\pm0.04$, $0.81^{+0.03}_{-0.04}$, $0.73\pm0.04$ for \textsc{NewHorizon}, \textsc{EAGLE}, \textsc{TNG} and \textsc{Magneticum} respectively, between $M_{\scriptstyle\star, \text{ 30 pkpc}}\sim10^{9.5}-10^{11.3}$ (\textsc{NewHorizon}), $10^{11.5}$ (\textsc{EAGLE}) and $10^{11.8}$ M$_{\scriptstyle \odot}$ (\textsc{TNG} and \textsc{Magneticum}) (Figure \ref{fig:stellar_mass_trend_explains_halo_mass}), showing the impact of stellar mass on the merger frequency of galaxies.
    \item Tidal feature fractions increase with increasing halo mass to a peak mass of $M_{200,\text{ crit}}\sim10^{12.7}$ M$_{\scriptstyle \odot}$ before declining with increasing halo mass (Figure \ref{fig:tidal_feature_fraction_halo_mass}).
    \item Central galaxies exhibit increasing tidal feature fractions with increasing halo mass, a trend that can be accounted for by the relationship between tidal feature fraction and galaxy stellar mass. We expect over half of the central galaxies in haloes with masses $\gtrsim M_{\scriptstyle 200, \rm crit}10^{12.7}$ M$_{\scriptstyle\odot}$ to have evidence of at least 1 tidal feature.
    \item Satellite galaxies tend to exhibit a declining tidal feature fraction for $M_{200, \rm crit}\gtrsim10^{13}$ M$_{\scriptstyle \odot}$, and this trend cannot be fully accounted for by the stellar masses of the galaxies alone, suggesting a potential additional environmental dependence (Figure \ref{fig:stellar_mass_trend_explains_halo_mass}).
    \item Comparison with observations indicate that our results are consistent with the tidal feature fractions that we would expect from LSST, complete to the tidal features with stellar masses $\gtrsim10^{6}$ M$_{\scriptstyle\odot}$ for \textsc{NewHorizon} that are disrupted to an area $\lesssim36$ pkpc$^2$ and $\gtrsim10^{8}$ M$_{\scriptstyle\odot}$ for \textsc{EAGLE}, \textsc{TNG} and \textsc{Magneticum} that are detectable even if disrupted to an area of $\sim2500$ pkpc$^2$, provided the light from the host in the same region is of a similar surface brightness or fainter than the tidal feature (Appendix \ref{app:detection_limits}). There is some indication from our comparisons with \citet{desmonsGalaxyMassAssembly2023} that shells might be detectable at lower host galaxy stellar masses in observations than in simulations.
\end{itemize}

We have produced predictions from cosmological simulations that are directly testable using LSST data, in particular exploration of trends of tidal feature fractions as a function of stellar and halo mass. Of particular interest will be using observed tidal features as a proxy for close encounters and mergers between galaxies, and determining how this fraction depends on halo mass to probe the galaxy assembly histories and the impact of the environment on them.

\section*{Acknowledgements}
We thank the anonymous referee for their thoughtful report that has improved the paper. We thank Annalisa Pillepich for helpful discussions regarding \textsc{IllustrisTNG} data. SB acknowledges funding support from the Australian Research Council through a Discovery Project DP190101943. This research includes computations using the computational cluster Katana supported by Research Technology Services at UNSW Sydney. This work was performed on the OzSTAR national facility at Swinburne University of Technology. The OzSTAR program receives funding in part from the Astronomy National Collaborative Research Infrastructure Strategy (NCRIS) allocation provided by the Australian Government, and from the Victorian Higher Education State Investment Fund (VHESIF) provided by the Victorian Government. The \textsc{NewHorizon} simulation was undertaken with HPC resources of CINES under the allocations  c2016047637, A0020407637 and A0070402192 by Genci, KSC-2017-G2-0003 by KISTI, and as a “Grand Challenge” project granted by GENCI on the AMD Rome extension of the Joliot Curie supercomputer at TGCC. A large data transfer was supported by KREONET which is managed and operated by KISTI. We acknowledge the Virgo Consortium for making their simulation data available. The eagle simulations were performed using the DiRAC-2 facility at Durham, managed by the ICC, and the PRACE facility Curie based in France at TGCC, CEA, Bruyères-le-Châtel. TNG100 was run on the HazelHen Cray XC40 system at the High Performance Computing Center Stuttgart as part of project GCS-ILLU of the Gauss Centre for Supercomputing (GCS). Ancillary and test runs of the IllustrisTNG project were also run on the Stampede supercomputer at TACC/XSEDE (allocation AST140063), at the Hydra and Draco supercomputers at the Max Planck Computing and Data Facility, and on the MIT/Harvard computing facilities supported by FAS and MIT MKI. The {\it Magneticum} simulations were performed at the Leibniz-Rechenzentrum with CPU time assigned to the Project {\it pr83li}. This work was supported by the Deutsche Forschungsgemeinschaft (DFG, German Research Foundation) under Germany's Excellence Strategy - EXC-2094 - 390783311. Parts of this research were supported by the Australian Research Council Centre of Excellence for All Sky Astrophysics in 3 Dimensions (ASTRO 3D), through project number CE170100013. 

Software: \textsc{Astropy} \citep{astropycollaborationAstropyCommunityPython2013,astropycollaborationAstropyProjectBuilding2018,astropycollaborationAstropyProjectSustaining2022}, \textsc{Numpy} \citep{harrisArrayProgrammingNumPy2020a}, \textsc{Pandas} \citep{mckinneyDataStructuresStatistical2010,thepandasdevelopmentteamPandasdevPandasPandas2023}, \textsc{Scipy} \citep{virtanenSciPyFundamentalAlgorithms2020}.

\section*{Data Availability}

\textsc{NewHorizon} data may be requested from \url{https://new.horizon-simulation.org/data.html}. \textsc{EAGLE} data used in this work are publically available at \url{http://icc.dur.ac.uk/Eagle/}. The \textsc{IllustrisTNG} data used in this work are publicly available at \url{http://www.tng-project.org}. \textsc{Magneticum} data are partially available at \url{https://c2papcosmosim.uc.lrz.de/} \citep{ragagninWebPortalHydrodynamical2017}, with
larger data sets on request.



\bibliographystyle{mnras}
\bibliography{references} 




\appendix

\section{Overview of subgrid physics models}
\label{sec:app_sims}
\subsection{NewHorizon}
\textsc{NewHorizon} uses the adaptive mesh refinement code (AMR) \textsc{RAMSES} \citep{teyssierCosmologicalHydrodynamicsAdaptive2002}. Star formation is set to occur in regions with hydrogen gas number density higher than $10$ cm$^{-3}$, following the Schmidt law: $\dot{\rho_{\star}}=\epsilon_\star\rho_{\rm{g}}/t_{\rm{ff}}$, which relates the star formation rate mass density, $\rho_{\star}$, to the gas mass density, $\rho_{\mathrm{g}}$, local free fall time of the gas, $t_{\rm{ff}}$ and lastly a varying star formation efficiency, $\epsilon_{\star}$ \citep{kimmImpactLymanAlpha2018,trebitschOBELISKSimulationGalaxies2021}. The star formation efficiency is computed following the prescriptions of \citet{hennebelleAnalyticalStarFormation2011}, which account for the turbulence of the gas. H and He gas are modelled using an equilibrium chemistry model with a homogeneous UV background. This gas is allowed to cool to $\simeq10^4$K through collisional ionisation, excitation, recombination, Bremsstrahlung, and Compton cooling. This gas, if metal enriched is allowed to further cool down to 0.1 K using the rates from \citet{dalgarnoHeatingIonizationHI1972} and \citet{sutherlandCoolingFunctionsLowDensity1993}.

The impact of supernovae (SNe) on the gas and therefore, star formation is another process below the resolution limit of cosmological simulations. \textsc{NewHorizon} models each SNe as an explosion releasing $10^{51}$ erg of energy into the surrounding baryons, with a minimum of 6 M$_{\scriptstyle \odot}$ needed in a star to go SN, they model a specific SNe rate of 0.03 M$_{\scriptstyle \odot}^{-1}$ using the mechanical SN feedback scheme described in \citet{kimmEscapeFractionIonizing2014} and \citet{kimmSimulatingStarFormation2015}.

\textsc{NewHorizon} implements two modes of active galactic nuclei (AGN) feedback, thermal and kinetic (radio and quasar), depending on the accretion rate of the supermassive black hole \citep[following][]{duboisJetregulatedCoolingCatastrophe2010,teyssierMassDistributionGalaxy2011}. AGN feedback can help remove gas from the galaxy and/or heat it up making it unsuitable for star formation.

\subsection{EAGLE}
\textsc{EAGLE} runs on the Lagrangian Smooth Particle Hydrodynamic (SPH) code \textsc{GADGET-3}. Star formation is implemented stochastically,  following the pressure law scheme of \citet{schayeRelationSchmidtKennicuttSchmidt2008}, implementing the observed Kennicut-Schmidt \citep{schmidtRateStarFormation1959,kennicuttGlobalSchmidtLaw1998} relation into the simulation. Star formation is set to occur in regions with a metallicity-dependent hydrogen mass density threshold, formulated by \citet{schayeStarFormationThresholds2004}, to account for the more efficient radiative transfer in metal-rich gas clouds. Stellar particles are assumed to follow a Chabrier IMF \citep{chabrierGalacticStellarSubstellar2003}, with stars spanning from 0.1-100 M$_{\scriptstyle \odot}$. \textsc{EAGLE} accounts for stellar winds, radiation and SNe feedback by implementing their collective feedback through thermal heating, distributing the energy produced at each timestep by a stellar particle to the neighbouring particles using the stochastic thermal feedback scheme of \citet{dallavecchiaSimulatingGalacticOutflows2012}.

The radiative cooling and heating rates of gas resolution elements at a given density, temperature and redshift are computed with the software \textsc{CLOUDY} \citep{ferlandCLOUDY90Numerical1998}. The gas is assumed to be optically thin, in an ionisation equilibrium and exposed to the cosmic microwave background and a spatially homogeneous UV/X-ray background that evolves over time \citep{haardtModellingUVXray2001,wiersmaEffectPhotoionizationCooling2009}. \textsc{EAGLE} implements one mode of AGN feedback through stochastic heating, following the prescriptions of \citet{dallavecchiaSimulatingGalacticOutflows2008} and \citet{boothCosmologicalSimulationsGrowth2009}.

\subsection{IllustrisTNG}
\textsc{TNG} used the moving mesh code \textsc{AREPO} \citep{springelPurSiMuove2010}, which differs from AMR code (e.g. \textsc{RAMSES}) and SPH code (e.g. \textsc{GADGET-3}) in that it uses an unstructured mesh defined by a Voronoi tessellation of a set of discrete points. This alternative methodology was formulated to address issues with AMR and SPH, which impact their accuracy in particular situations (e.g. suppression of fluid instabilities in SPH and lack of Galilean invariance and overmixing present in AMR codes).

\textsc{TNG} models star formation stochastically, treating the star formation and pressurisation of a multi-phase interstellar medium following \citet{springelCosmologicalSmoothedParticle2003}. In the model by \citet{springelCosmologicalSmoothedParticle2003}, cold gas above a density threshold of $0.1$ H cm$^{-3}$ forms star particles following the empirically defined Kennicutt-Schmidt relation \citep{schmidtRateStarFormation1959,kennicuttGlobalSchmidtLaw1998} and the Chabrier IMF \citep{chabrierGalacticStellarSubstellar2003}. Under the \citet{springelCosmologicalSmoothedParticle2003} prescription, SNe pressurises the gas and can lead to enhanced star formation. The stellar winds produced by star formation can carry and inject kinetic energy into the surrounding gas. They are modelled as described in \citet{vogelsbergerModelCosmologicalSimulations2013} and \citet{pillepichSimulatingGalaxyFormation2018}.

The radiative cooling of gas is modelled accounting for its metal enrichment \citep[following][]{wiersmaEffectPhotoionizationCooling2009},  and a time-evolving homogeneous UV background with self-shielding corrections in dense interstellar medium \citep[following][]{katzGalaxiesGasCold1992,faucher-giguereNewCalculationIonizing2009}. Following the models of \citet{vogelsbergerModelCosmologicalSimulations2013}, radiative cooling is further influenced by the nearby radiation fields of AGN.

\textsc{TNG} implements subgrid AGN feedback with a radio and quasar mode. For accretion rates below 5\% of the Eddington limit, the radio-mode feedback is active, injecting bursty thermal energy into a $\sim50$ pc bubble displaced away from the host galaxy \citep{sijackiUnifiedModelAGN2007}. For higher accretion rates the quasar mode is active, injecting thermal energy continuously into the adjacent gas \citep{springelModellingFeedbackStars2005,dimatteoEnergyInputQuasars2005}  The details of all subgrid physics can be found in \citet{weinbergerSimulatingGalaxyFormation2017,pillepichSimulatingGalaxyFormation2018} and \citet{nelsonIllustrisTNGSimulationsPublic2019}.

\subsection{Magneticum}
\textsc{Magneticum Pathfinder} simulations are a suite of cosmological hydrodynamical simulations, ranging in box size from $25.6^3$ to 3818$^3$ cMpc$^3$. We use the \textsc{Box4-uhr} simulation, which models the physics of dark matter and baryons in a $68^3$ cMpc$^3$ box. \textsc{Magneticum Pathfinder} has been performed with a modified version of the Lagrangian Smooth Particle Hydrodynamic (SPH) code \textsc{GADGET-3}, with modifications on the viscosity and treatment of the kernel functions following \citet{dolagTurbulentGasMotions2005}, \citet{donnertRiseFallRadio2013}, and \citet{beckImprovedSPHScheme2016}. Star formation and the stellar wind-driven kinetic feedback are modelled following \citet{springelCosmologicalSmoothedParticle2003}. Each star particle represents a stellar population following a Chabrier IMF \citep{chabrierGalacticStellarSubstellar2003} and each gas particle can form up to 4 stars. Star formation and metal enrichment from supernova feedback and Asymptotic Red Giant Branch stars is modelled following the \citet{springelCosmologicalSmoothedParticle2003} as well as the local metallicity dependent processes \citep{wiersmaEffectPhotoionizationCooling2009,dolagDistributionEvolutionMetals2017}. The radiative cooling and heating rates of gas resolution elements at a given density, temperature and redshift are computed with the software CLOUDY \citep{ferlandCLOUDY90Numerical1998}. The radiative cooling accounts for a UV/X-ray background following the prescriptions of \citet{haardtModellingUVXray2001} and \citet{wiersmaEffectPhotoionizationCooling2009}.

The subgrid physics treatment of supermassive black holes and their AGN feedback is implemented as described by \citet{fabjanSimulatingEffectActive2010} and \citet{hirschmannCosmologicalSimulationsBlack2014}. The implemented black holes feedback scheme accounts for a transition from quasar to radio mode according to \citep{sijackiUnifiedModelAGN2007}. Note that the black holes in this simulation are not pinned to the potential minimum. Thermal conduction is implemented according to \citet{dolagThermalConductionSimulated2004} but following \citet{arthAnisotropicThermalConduction2017}, with 1/20 of the classical Spitzer value \citep{spitzerPhysicsFullyIonized1962}. The details of all subgrid physics are described by \citet{tekluConnectingAngularMomentum2015} and \citet{hirschmannCosmologicalSimulationsBlack2014}.

\section{Tidal feature detection limits}
\label{app:detection_limits}
\begin{figure}
    \centering
    \includegraphics[width=\linewidth]{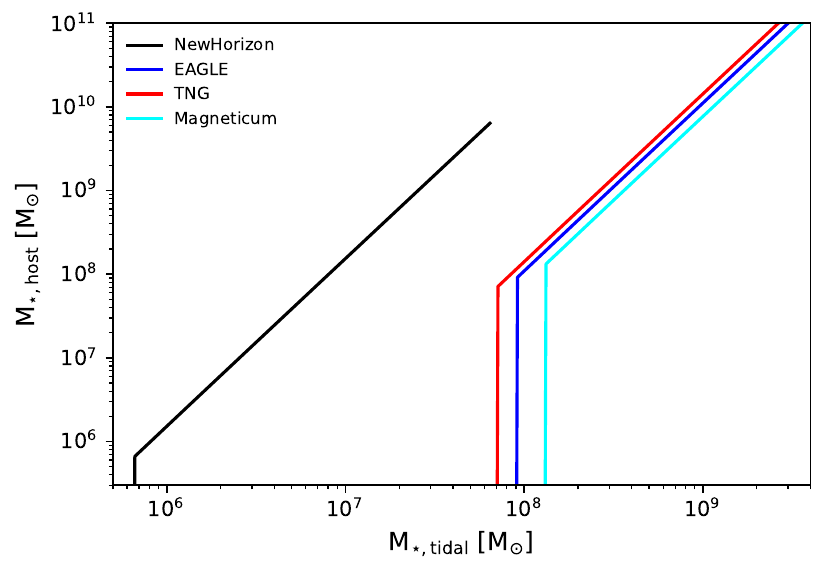}
    \caption{The maximum stellar mass from the host galaxy that in the same region as the tidal feature without making it undetectable as a function of the tidal feature stellar mass. For a tidal feature to be detectable it has SNR$>5$.}
    \label{fig:mtidal_v_mgal}
\end{figure}

\begin{figure*}
    \centering
    \includegraphics[width=\linewidth]{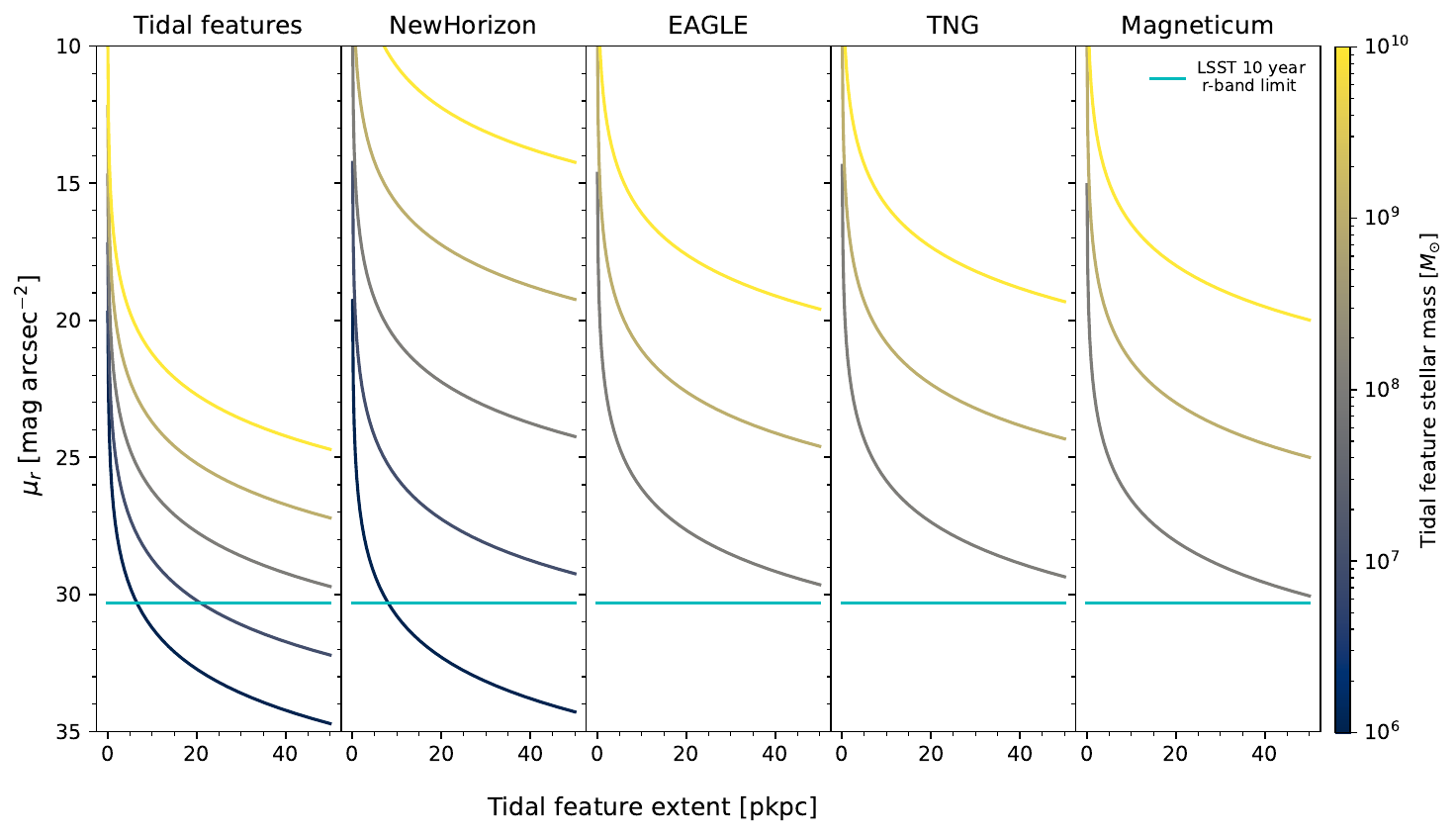}
    \caption{The $r$-band surface brightness of the model tidal features as a function of their radii and the corresponding maximum possible surface brightness of the light from the host galaxy within the tidal features area such that the tidal feature is still detectable. The tidal features panel shows the surface brightness of each tidal feature as a function of the radius of the circle the light is spread over. The remaining panels show the maximum surface brightness of the light from the host that can be within the same region as the tidal feature without making it undetectable for each simulation. The colours correspond to the stellar masses of the tidal features, $10^6$, $10^7$, $10^8$, $10^9$ and $10^{10}$ M$_{\scriptstyle\odot}$. The LSST 10 year $r$-band, $3\sigma$, $10''\times10''$, limiting surface brightness is shown in Rubin Turquoise.}
    \label{fig:feature_and_host_sb}
\end{figure*}

The stellar mass resolutions of our simulations as well as the surface brightness limits of our mock images limits the minimum stellar masses of the tidal features we can resolve. Through some simple modelling we investigate the tidal feature detection limits for each of our simulations. Following \citet{martinPreparingLowSurface2022} we calculate the signal-to-noise ratio for a tidal feature against its background galaxy and take a SNR$_{\mathrm{tidal}}>5$ as the range for a detectable tidal feature.
\begin{equation}
    \mathrm{SNR}_{\rm tidal}=\frac{n_{\rm tidal}}{\sigma_{\rm tidal}} = \frac{n_{\rm tidal}}{\sqrt{2n_{\rm host}+n_{\rm tidal}}}
\end{equation}
Here, $n_{\rm tidal}$ is the number of particles that comprise a tidal feature, $n_{\rm host}$ is the number of particles from the host in the same region as the tidal feature. Using this equation and the simulation mass resolutions (Table \ref{tab:simulations}) we determined for each of our simulations, given a tidal feature of a specific stellar mass, what was the maximum amount of stellar mass from the host that could be in the same region without making the tidal feature undetectable. The results of this calculation are shown in Figure \ref{fig:mtidal_v_mgal}, where we show the maximum host stellar mass contamination in the same region as the tidal feature as a function of tidal feature stellar mass. We can see that each simulation has a drop off corresponding to where its mass resolution is limiting its ability to resolve tidal features with SNR$>5$ without there being virtually no contamination from the host galaxy. From this calculation we can predict that \textsc{NewHorizon} is able to resolve tidal features that are $\gtrsim2$ orders of magnitude less massive than in the other three simulations. \textsc{NewHorizon} should also be able to resolve tidal features with much greater contamination from the host galaxy when compared to the other three simulations.

To investigate the minimum mass galaxy that can be tidally disrupted and detected in our mock images, we must construct a model for the surface brightnesses of tidal features stemming from different stellar mass hosts. We build a simple model assuming the following:
\begin{itemize}
    \item the mass-to-light ratio for the stellar particles is 1,
    \item the light from a tidal feature is homogeneously distributed across a square region.
\end{itemize}

While mass-to-light ratios vary as a function of galaxy morphology \citep[e.g.][]{dezeeuwSAURONProjectII2002,emsellemSAURONProjectIII2004,cappellariSAURONProjectOrbital2007} and could vary significantly across a galaxy \citep[e.g.][]{mehrganDynamicalStellarMasstolight2024}, 1 is a physical mass-to-light ratio and eases computation substantially. The homogeneous distribution of light across a square area is a simplifying assumption that allows us to compute the surface brightness of tidal features without simulating features of varying light distributions and regions. If the mass-to-light ratio for the tidally disrupted galaxy is above 1 or the distribution of the light is inhomogenous resulting in some regions that are brighter than the average brightness of the tidal feature, the resultant tidal features would be detectable despite having lower stellar mass than our estimates.

We convert from the physical surface brightness of the stellar particles in L$_{\scriptstyle\odot}\text{ pc}^{-2}$ to LSST r-band surface brightness in mag arcsec$^{-2}$ using the following relationship:
\begin{equation}
    \mu_{r}=-2.5\log_{10}(\mathrm{SB}(\mathrm{L}_{\scriptstyle\odot}\mathrm{\:pc}^{-2}))+M_{\scriptstyle\odot,\mathrm{ r}} + 5\log_{10}\bigg(\frac{648 000}{10\pi}\bigg)
\end{equation}
where we use the estimated absolute AB magnitude of the Sun in the LSST $r$-band, $M_{\scriptstyle\odot,\mathrm{ r}}=4.64$ \citep{willmerAbsoluteMagnitudeSun2018}.

The left most panel in Figure \ref{fig:feature_and_host_sb} shows the surface brightness with tidal feature extent for tidal features of stellar masses $10^{6}$, $10^{7}$, $10^{8}$, $10^{9}$ and $10^{10}$ M$_{\scriptstyle\odot}$. The surface brightness profiles in remaining panels of Figure \ref{fig:feature_and_host_sb} indicate the maximum allowed host brightness within the same area to still allow the tidal feature to be resolved against the host background contamination. We compute these assuming a minimum SNR$_{\rm tidal}>5$ and again assuming the host light is distributed evenly in the background. Our results highlight that even tidal features from galaxies that are 10$^{6}$ M$_{\odot}$ are above the mock image surface brightness limits provided their radius remains within $\sim6$ pkpc. However, only \textsc{NewHorizon} has the sufficient resolution to resolve them so long as the light from the host galaxy in the same region as the tidal feature is of a similar brightness or fainter. The surface brightness profile for the maximum allowed host contamination indicates that the these features are only detectable if there is a similar amount of light from the host in this region. From Figure \ref{fig:mtidal_v_mgal} we can estimate that \textsc{NewHorizon} will resolve a $10^{6}$ M$_{\scriptstyle\odot}$ feature within the survey detection limits if there is $\lesssim1.5\times10^{6}$ M$_{\scriptstyle\odot}$ of stellar mass from the host within the same region, provided the feature is no more extended than $\sim6$ pkpc. For the larger cosmological simulations, we can detect tidal features of stellar mass $\gtrsim10^{8}$ M$_{\scriptstyle\odot}$ even if they have been disrupted over extents $\gtrsim50$ pkpc, so long as the host stellar contamination is $\lesssim1.5\times10^{8}$ M$_{\scriptstyle\odot}$ within the same region. As the tidal features become more massive than these lower limits, the maximum surface brightness of the host galaxy within the same region as the tidal feature increases for each of the simulations.


\bsp	
\label{lastpage}
\end{document}